\input{epsf}
\newcommand{\et}{\hspace{-0.08in}{\bf .}\hspace{0.1in}}
\newcommand{\BOX}{\hbox {$\sqcap$ \kern -1em $\sqcup$}}

\newcounter{letter} 
\newenvironment{alphalist}{
\begin{list}{{\normalshape(\alph{letter})}}{\usecounter{letter}}
}{\end{list}}

\newcommand{\so}{{\frak so}}
\newcommand{\R}{{\Bbb R}}
\newcommand{\C}{{\Bbb C}}
\newcommand{\Z}{{\Bbb Z}}

\renewcommand{\to}{\rightarrow}
\newcommand{\tensor}{\otimes}
\newcommand{\maps}{\colon}

\newcommand{\iso}{\cong}
\newcommand{\we}{\wedge}

\renewcommand{\H}{{\cal H}}
\newcommand{\F}{{\cal F}}
\newcommand{\A}{{\cal A}}
\newcommand{\T}{{\cal T}}
\newcommand{\X}{{\cal X}}
\newcommand{\G}{{\cal G}}

\newcommand{\SL}{{\rm SL}}
\newcommand{\SU}{{\rm SU}}

\newcommand{\SO}{{\rm SO}}

\newcommand{\Spin}{{\rm Spin}}

\newcommand{\Inv}{{\rm Inv}}
\newcommand{\tr}{{\rm tr}}

\newtheorem{thm}{Theorem}    

\newtheorem{prop}[thm]{Proposition}
\newtheorem{defn}[thm]{Definition}

        \newcommand{\be}{\begin{equation}}
        \newcommand{\ee}{\end{equation}}
        \newcommand{\ba}{\begin{eqnarray}}
        \newcommand{\ea}{\end{eqnarray}}
        \newcommand{\ban}{\begin{eqnarray*}}
        \newcommand{\ean}{\end{eqnarray*}}
        \newcommand{\barr}{\begin{array}}
        \newcommand{\earr}{\end{array}}

\documentstyle[12pt,amsfonts]{article}

\begin{document}

      \begin{center}
      {\bf Spin Foam Models \\}
      \vspace{0.5cm}
      {\em John C.\ Baez\\}
      \vspace{0.3cm}
      {\small Department of Mathematics, University of California\\ 
      Riverside, California 92521 \\
      USA\\ }
      \vspace{0.3cm}
      {\small email: baez@math.ucr.edu\\}
      \vspace{0.3cm}
      {\small September 19, 1997 \\ }
      \end{center}

\begin{abstract}  
\noindent
While the use of spin networks has greatly improved our understanding of
the {\it kinematical} aspects of quantum gravity, the {\it dynamical}
aspects remain obscure.  To address this problem, we define the concept
of a `spin foam' going from one spin network to another.  Just as a spin
network is a graph with edges labeled by representations and vertices
labeled by intertwining operators, a spin foam is a 2-dimensional
complex with faces labeled by representations and edges labeled by
intertwining operators.  Spin foams arise naturally as
higher-dimensional analogs of Feynman diagrams in quantum gravity and
other gauge theories in the continuum, as well as in lattice gauge
theory.  When formulated as a `spin foam model', such a theory consists
of a rule for computing amplitudes from spin foam vertices, faces, and
edges.  The product of these amplitudes gives the amplitude for the spin
foam, and the transition amplitude between spin networks is given as a
sum over spin foams.  After reviewing how spin networks describe
`quantum 3-geometries', we describe how spin foams describe `quantum
4-geometries'.  We conclude by presenting a spin foam model of
4-dimensional Euclidean quantum gravity, closely related to the state sum 
model of Barrett and Crane, but not assuming the presence of an underlying
spacetime manifold. 
\end{abstract}

\section*{Introduction}

Thanks in part to the introduction of spin network techniques \cite{RS},
we now have a mathematically rigorous and intuitively compelling picture
of the {\it kinematical} aspects of what used to be called the `loop
representation' of quantum gravity \cite{RS0}.  Indeed, since spin
networks form a very convenient basis of kinematical states \cite{B3},
they have largely replaced collections of loops as our basic picture of
`quantum 3-geometries'.  Unfortunately, the {\it dynamical} aspects of
quantum gravity remain much less well understood in this approach.  Over
and above the many technical problems, the crucial fact is that we lack
a simple picture of `quantum 4-geometries'.  However, recent work by
various authors \cite{I,R1,RR} suggests such a picture, which we
formalize and explore here using the notion of a `spin foam'.

The basic idea is simple.  Just as a spin network is a graph with edges
labeled by spins and vertices labeled by intertwining operators, a
`spin foam' is a 2-dimensional piecewise linear cell complex ---
roughly, a finite collection of polygons attached to each other along
their edges --- with faces labeled by spins and edges labeled by
intertwining operators.  
As with spin networks, we may think of spin foams either abstractly or
embedded in spacetime.  Either way, a generic slice of a spin foam `at
fixed time' gives a spin network.  Edges of this spin network come from
faces of the spin foam, while vertices of the spin network come from
edges of the spin foam.  As we move the slice `forwards in time', the
spin network changes topology only when the slice passes a vertex of the
spin foam.

In their joint work, Reisenberger and Rovelli \cite{RR} arrive at spin foams
through the study of quantum gravity on a manifold of the form $\R
\times S$ for some 3-manifold $S$ representing space.   They begin with
the Hamiltonian constraint $H$ with constant lapse function  as an
operator on the space of kinematical states.  Of course, the actual form
of this operator (if indeed it exists) is currently a matter of 
controversy: Thiemann has proposed a formula \cite{T2}, but it is far
from universally accepted.  Luckily, Reisenberger and Rovelli's argument
depends only on some general assumptions as to the nature of the
operator $H$.   In particular, they assume it generates a one-parameter
group $\exp(-itH)$.  Since $H$ is a Hamiltonian constraint rather than a
Hamiltonian, the physical interpretation of $\exp(-itH)$ is somewhat
problematic.  Nonetheless, they argue that it is an object of
substantial interest, encoding much of the dynamics of quantum gravity.

Under some assumptions on the form of the Hamiltonian constraint,
Reisenberger and Rovelli are able to compute the transition amplitude
$\langle \Psi, \exp(-itH) \Phi \rangle$ as a formal power series in $t$
for any spin network states $\Psi$ and $\Phi$.    The coefficient of
$t^n$ in this power series is a sum over certain equivalence classes of
spin foams embedded in spacetime with $\Psi$ as their initial slice,
$\Phi$ as their final slice, and a total of $n$ foam vertices.  Each
spin foam contributes an amplitude given by a product over its vertices
of certain `spin foam vertex amplitudes'.  In models with `crossing
symmetry', the spin foam vertex amplitude can be computed using only the
spin network obtained by intersecting the spin foam with a small sphere
about the vertex, and result depends only on the isotopy type of this
spin network.  
This is to be expected in Euclidean quantum gravity, since it amounts to
saying that applying a 4-dimensional rotation to a spin foam vertex does
not affect its amplitude.  In what follows we only consider models with
crossing symmetry.  (Unfortunately, Thiemann's formula for the
Hamiltonian constraint lacks crossing symmetry.)

In this context, spin foams play a role much like
that of Feynman diagrams (see Figure 1). 
In standard quantum field theory we compute transition amplitudes as
sums or integrals over graphs with edges labeled by irreducible unitary
representations of the relevant symmetry group.  Typically this group is
the product of the Poincar\'e group and some internal symmetry group, so
the edges are labeled by momenta, spins, and certain internal quantum
numbers.  To compute the transition amplitude from one basis state to
another, we sum over graphs going from one set of points labeled by
representations (and vectors lying in these representations) to some
other such set.   The contribution of any graph to the amplitude is
given by a product of amplitudes associated to its vertices and edges.  
Each vertex amplitude depends only on the representations labeling the
incident edges, while each edge amplitude, or propagator, depends only
on the label of the edge itself.   The propagators are usually computed
using a free theory about which we are doing a perturbative expansion,
while the vertices represent interactions.  

\vbox{
\medskip
\centerline{\epsfysize=2in\epsfbox{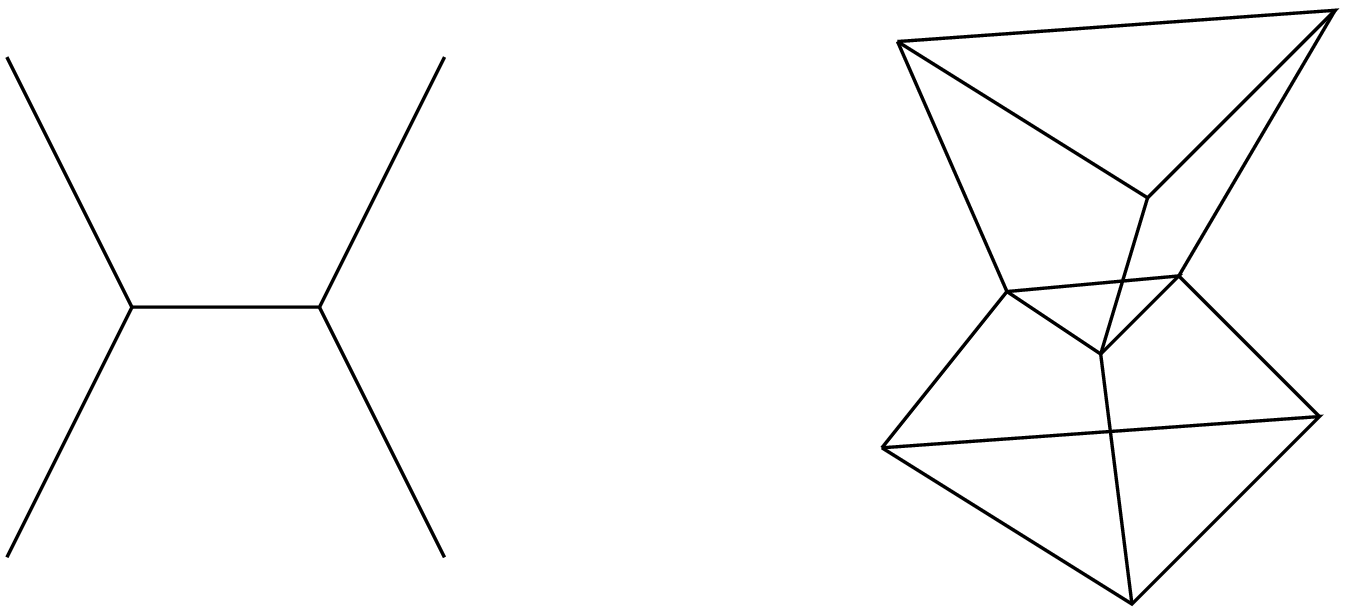}}
\bigskip
\centerline{1.  Feynman diagram versus spin foam}
\medskip
}

Similarly, we can consider `spin foam models' for an arbitrary symmetry
group $G$, generalizing the case considered by Reisenberger and Rovelli,
in which $G = \SU(2)$.  In this more general context a spin network is
defined as a graph with oriented edges labeled by irreducible unitary
representations of $G$ and vertices labeled by intertwining operators
from the tensor product of the representations labeling `incoming'
incident edges to the tensor product of representations labeling
`outgoing' edges.   By analogy, we define a spin foam to be a
2-dimensional piecewise linear cell complex with oriented faces labeled
by irreducible unitary representations of $G$ and oriented edges labeled
by intertwining operators.

In such a model we compute the transition amplitude between two spin 
networks as a sum over spin foams going from the first spin network to
the second.  Each spin foam contributes to the amplitude an amount given
by a product of amplitudes associated to its vertices, edges, and faces.
In a model with crossing symmetry, the amplitude of any vertex depends
only on the isotopy class of the spin network obtained by intersecting
the spin foam with a small sphere centered at the vertex.
The amplitude of any edge depends only on the intertwining operator
labeling that edge and the representations labeling the incident faces. 
The amplitude of any face depends only on the representation labeling
that face.  By analogy with Feynman diagrams, the edge and face
amplitudes can be thought of as `propagators'.  Spin foam vertices can
be thought of as `interactions', and the vertex amplitudes characterize
the nontrivial dynamics of the theory.  

Reisenberger and Rovelli's approach to spin foams starting from
canonical quantum gravity raises a number of problems: the
well-definedness of the Hamiltonian constraint, the physical meaning of
the one-parameter group $\exp(-itH)$, the mathematical sense in which it
can be approximated by a power series in $t$, and so on.  It is
reassuring therefore that spin foams also arise naturally in lattice
gauge theory, where these problems are absent.  The basic idea, due to
Iwasaki \cite{I} and Reisenberger \cite{R1}, is that a path integral in
lattice gauge theory can be represented as a sum over spin foams living
in the lattice.  The advantage of the lattice context is that it avoids
some of the deep issues of analysis involved in the usual path-integral
approach to quantum gravity.  One can then address these issues by
studying the limit in which the lattice is subdivided ever more finely.

Finally, in addition to canonical quantum gravity and lattice quantum
gravity, it is worth comparing another approach to quantum gravity in
which transition amplitudes  can be computed using 2-dimensional
analogs of Feynman diagrams: perturbative string theory.  There are two
major differences between spin foam models and perturbative string
theory.  First, perturbative string theory considers only 2-manifolds
with boundary mapped into spacetime, not more general piecewise linear
cell complexes.  Second, transition amplitudes in perturbative string
theory are computed with the aid of a fixed metric on the ambient
spacetime, while spin foam models make no reference to such a metric. 
These two facts are closely related.  Since strings do not interact when
they intersect in perturbative string theory, interactions arise
primarily from the coupling of the string to the fixed metric on the
ambient spacetime.  In spin foam models, interactions occur where spin
foam faces meet at vertices, so there is no need for a fixed background
metric.   

Interestingly, recent work on nonperturbative string theory suggests
that strings and other higher-dimensional membranes interact where they
intersect.  Historically speaking, string theory and the loop
representation of quantum gravity are twins separated at birth.  Both
arose from the failure of string theory as a viable fundamental theory
of hadrons and the resulting attempts to reconcile its phenomenological
successes with quantum chromodynamics.  The two theories went very
different ways, but they are still related.  The author has already
argued that the loop representation of quantum gravity can be viewed as
a novel sort of background-free string theory \cite{B1}, and shown that
we may naturally associate a Wess-Zumino-Witten model to the surfaces
appearing in the spacetime formulation of the loop representation of
quantum gravity \cite{B5}.  One reason for studying spin foams is to
understand this better.

The plan of the paper is as follows.  In Section 1 we give a precise
definition of spin foams.   It turns out that there is a category whose
objects are spin networks and whose morphisms are spin foams.  The basic
idea is that a morphism $F \maps \Psi \to \Psi'$  in this category is a
spin foam $F$ going from the spin network $\Psi$ to the spin network
$\Psi'$.  We can glue a spin foam $F \maps \Psi \to \Psi'$ to a spin
foam $F' \maps \Psi' \to \Psi''$ and obtain a spin foam $FF' \maps \Psi
\to \Psi''$.  This sets up a useful analogy: `spin foams are to
operators as spin networks are to states'.  

In addition to these `abstract' spin foams one can consider spin foams 
embedded in a spacetime manifold.  Various sorts of embeddings are
interesting, with continuous, piecewise-linear, smooth and real-analytic
ones being the most important.  To keep this paper from becoming too
long we discuss only one case here: piecewise-linearly embedded spin
foams.  The advantages of piecewise-linearly embedded spin networks have
been pointed out by Zapata \cite{Z}, the main one being the
simplicity of a purely combinatorial approach.  The same advantages
apply when studying spin foams, but we emphasize that all options should
be explored at this point.  In Section 2 we sketch the role of
piecewise-linearly embedded spin foams in lattice gauge theory.  

In Section 3 we review the interpretation of spin networks as quantum
3-geometries.  We work with $\SU(2)$ spin networks embedded in the
dual 1-skeleton of a triangulated 3-manifold $S$.  In other words, we
consider only spin networks having one 4-valent vertex at the center
of each tetrahedron in $S$ and one edge intersecting each triangle in
$S$.  For $\SU(2)$, irreducible representations are described by spins
$j = 0, {1\over 2}, 1,\dots$, and an intertwining operator
at a 4-valent vertex can also be described by a spin.  Thus we may
work in the `dual picture' and think of a spin network state as a
labeling of all the triangles and tetrahedra in $S$ by spins
\cite{HP,Im, Markopoulou,R}.  Thanks in part to Barbieri's work on the
quantum tetrahedron \cite{Barb}, the Hilbert space spanned by these
states can be interpreted as the space of `quantum 3-geometries' of
the triangulated 3-manifold $S$.  In particular, one can define an
`area operator' for each triangle and a `volume operator' for each
tetrahedron.  These operators are closely related to the area and
volume operators first introduced by Rovelli and Smolin \cite{RS2}, as
well as the slightly different ones studied by Ashtekar, Lewandowski
and collaborators \cite{AL2.5,AL3,AL4,ALMMT,L}.

In Section 4 we turn to the interpretation of spin foams as quantum
4-geometries.  Reisenberger and Rovelli \cite{RR} have already
considered this issue for real-analytically embedded spin foams with
gauge group $\SU(2)$.  Here we use $\Spin(4)$ spin foams embedded in the
dual 2-skeleton of a triangulated 4-manifold $M$.  Such spin foams have
one vertex centered at each 4-simplex in $M$, one edge intersecting each
tetrahedron, and one face intersecting each triangle.  Since
\[            \Spin(4) \iso \SU(2) \times \SU(2), \]
we can describe such a spin foam in the dual picture by labeling each
triangle and each tetrahedron in $M$ by a pair of spins.  Restricting
attention to the `left-handed' copy of $\SU(2)$, such a spin foam
determines an $\SU(2)$ spin network in any 3-dimensional submanifold $S
\subseteq M$ compatible with the triangulation.  Thus, by the results
described in the previous section, the spin foam endows each such
submanifold with a quantum 3-geometry.  However, for these quantum
3-geometries to fit together to form a sensible quantum 4-geometry, it
appears that certain constraints must hold.  

Following the ideas of Barrett and Crane \cite{BC}, we arrive at these
constraints through a study of the `quantum 4-simplex'.  Here we take
$S$ to be the boundary of a single 4-simplex in $M$.  A spin foam in $M$
gives each of the ten triangles in $S$ an area and each of the five
tetrahedra in $S$ a volume.  However, the geometry of a 4-simplex
affinely embedded in Euclidean $\R^4$ is determined by only ten numbers,
e.g., the lengths of its edges.  This suggests that some constraints must
hold for $S$ to be the boundary of a `flat' 4-simplex.  We wish the
4-simplices in $M$ to be flat because we want a picture similar to that
of the Regge calculus, where spacetime is pieced together from flat
4-simplices, and curvature is concentrated along their boundaries
\cite{Regge}.

Interestingly, these constraints also arise naturally from the
relationship between general relativity and $BF$ theory, a 4-dimensional
topological field theory \cite{B4.5,B5}.  Let us recall how this works
in the case of signature ++++.  Suppose we have a smooth 4-manifold $M$
and a vector bundle $\T \to M$ isomorphic to the tangent bundle and
equipped with an orientation and positive-definite metric.  There is a
formulation of general relativity in which the basic fields are a
metric-preserving connection $A$ on $\T$ and a $\T$-valued 1-form $e$,
usually called the `cotetrad' field.  The action is given by
\[      \int_M \tr(e \we e \we F),  \]
where $F$ is the curvature of $A$, the wedge product of $\Lambda
\T$-valued forms is taken in the obvious way, and the map $\tr \maps
\Lambda^4 \T \to \R$ is defined using the metric and orientation on
$\T$.  The equations of motion are
\[     d_A (e \we e) = 0, \qquad  e \we F = 0, \]
where $d_A$ denotes the exterior covariant derivative.  When $e$ is
nondegenerate these equations are equivalent to the vacuum Einstein
equations.

On the other hand, one can modify this formulation, working not with the
cotetrad field $e$ but with the field $E = e \we e$.  First, note that
if one takes $E$ to be an arbitrary $\Lambda^2 \T$-valued 2-form and
uses the action
\[   \int_M \tr(E \we F)  \]
one obtains not general relativity but $BF$ theory,
whose equations of motion are simply
\[       d_A E = 0, \qquad F = 0 .\]
The reason is that not every such $E$ field can be written as $e \we e$
for some cotetrad field $e$.  For this to hold, it must satisfy some extra
constraints.  If one finds stationary points of the above action subject
to these constraints, one obtains equations equivalent to the vacuum
Einstein equations, at least when $E$ is nondegenerate.

Now, doing general relativity with the cotetrad field $e$ is very much
like describing 4-simplices using vectors for edges, while doing general
relativity with the $E$ field is very much like describing 4-simplices
using bivectors for faces.  Suppose we have a 4-simplex affinely embedded
in $\R^4$.  We can number its vertices $0,1,2,3,4$, and translate it so
that the vertex 0 is located at the origin.  Then one way to describe
its geometry is by the positions $e_1,e_2,e_3,e_4$ of the other four
vertices.  Another way is to use the bivectors
\[             E_{ab} = e_a \we e_b  \in \Lambda^2 \R^4 \]
corresponding to the six triangular faces with 0 as one of their vertices.
However, not every collection of bivectors $E_{ab}$ comes
from a 4-simplex this way.  In addition to the obvious skew-symmetry
$E_{ab} = -E_{ba}$, some extra constraints must hold.  They have
exactly the same form as the extra constraints that we need to obtain 
general relativity from $BF$ theory.  

Barrett and Crane showed that these constraints take a particularly nice
form if we describe them in terms of bivectors corresponding to all ten
triangular faces of the 4-simplex.  Quantizing these constraints, one
obtains conditions that a $\Spin(4)$ spin foam must satisfy to describe
a quantum 4-geometry built from flat quantum 4-simplices.  Interestingly,
these conditions also guarantee that the quantum geometries for
3-dimensional submanifolds obtained using the right-handed copy of
$\SU(2)$ agree with those coming from the left-handed copy.

In Section 5 we briefly describe two spin foam models for Euclidean
quantum gravity in four dimensions: the recently proposed state sum
model of Barrett and Crane (which may dually be formulated as a spin
foam model), and a closely related model in which no background
spacetime manifold is assumed.  This section is only intended as a
sketch of a future, more careful investigation of specific spin foam
models.   

\section{Spin foams}  \label{spin.foams}

A spin network is a graph with edges labeled by representations and
vertices labeled by intertwining operators, or `intertwiners' for
short.  Spin foams are like spin networks, but with everything one
dimension higher.  There are many choices of what we might take as 
higher-dimensional analogues of graphs.  In decreasing order of
generality, some well-known possibilities include CW complexes,
piecewise linear cell complexes, combinatorial CW complexes, and
simplicial complexes \cite{FP,HMS,H,RoSa}.  We shall work with piecewise
linear cell complexes,  but it is good to keep in mind that there are
other alternatives with their own advantages.  Also, while one tends to
think of 2-dimensional topology as a rather trivial business because
2-manifolds are easy to classify, it is worth noting that
there are many deep unsolved problems concerning the various sorts of
2-dimensional complexes, such as Whitehead's asphericity question, the
Andrews-Curtis conjecture, and Zeeman's conjecture.  In working with
spin foams, therefore, we should keep in mind that there is a lot of
pre-existing mathematical infrastructure to draw upon, but also some
deep questions yet to be understood.  The book by Hog-Angeloni, Metzler
and Sieradski serves as an excellent overview to all this \cite{HMS}.

In what follows we work with piecewise linear cell complexes as defined
by Rourke and Sanderson \cite{RoSa}.   The precise definition can be
found in Appendix A, but we provide a rough sketch here.  The basic
notion involved is that of a `cell'.  A $0$-cell is simply a point, a
$1$-cell is a closed interval, a $2$-cell is a polygon, and so on.  Any
cell has certain lower-dimensional cells (together with itself) as
`faces', and we write $Y \le X$ if $Y$ is a face of $X$.  A piecewise
linear cell complex, or `complex' for short, is a collection $\kappa$ of
cells in $\R^n$ such that the intersection of any two cells in $\kappa$
is again in $\kappa$, and the face of any cell in $\kappa$ is again in
$\kappa$.  In what follows we consider only finite complexes.  We write
$|\kappa|$ for the subset of $\R^n$ consisting of the union of the cells
in $\kappa$.

We often need to equip the cells of a complex with orientations. We say
a $k$-cell $X$ is {\it oriented} if $X$ minus the union of its proper
faces, thought of as a $k$-dimensional manifold, is equipped with an
orientation, and we say a complex is {\it oriented} if each $k$-cell
with $k \ge 1$ is oriented.  Here we implicitly equip each $0$-cell with
its positive orientation. Note that we do not require any consistency
between the orientations of the different cells.

To define spin networks we need some notation concerning 1-dimensional
oriented complexes.  Such a complex has a set $V$ of 0-cells or
{\it vertices}, and also a set $E$ of oriented 1-cells or {\it
edges}.  The orientation on each edge $e \in E$ picks out one of its
endpoints as its {\it source} $s(e) \in V$ and the other as its {\it
target} $t(e) \in V$.  If $v$ is the source of $e$ we say $e$ is {\it
outgoing} from $v$, while if $v$ is the target of $e$ we say that $e$ is
{\it incoming} to $v$.

As noted in Smolin's excellent review article \cite{S}, one of the
virtues of the concept of `spin network' is its versatility.  As a
result, no single definition easily captures the whole concept.  One can
study spin networks either abstractly or embedded in a topological,
piecewise linear, smooth or real-analytic manifold.  In addition, one
can label the spin network edges by representations of either groups or
quantum groups.  Here we define spin networks for any compact
group $G$, which we think of as the gauge group of the physical
theory in question \cite{B3}.  We pick a set of irreducible unitary
continuous representations of $G$, one from each equivalence class, to
use as labels.  Each representation $\rho$ in this set has a dual
$\rho^\ast$ in this set.  Henceforth when we speak of an irreducible
representation of $G$ we always mean one in this set.

\begin{defn}\et A {\it spin network} $\Psi$ is a triple
$(\gamma,\rho,\iota)$ consisting of:
\begin{enumerate}
\item a 1-dimensional oriented complex $\gamma$, 
\item a labeling $\rho$ of each edge $e$ of $\gamma$ by an
irreducible representation $\rho_e$ of $G$,
\item a labeling $\iota$ of each vertex $v$ of $\gamma$ by an intertwiner
\[       \iota_v \maps \rho_{e_1} \tensor \cdots \tensor \rho_{e_n} 
\to \rho_{e'_1} \tensor \cdots \tensor \rho_{e'_m}  \]
where $e_1,\dots,e_n$ are the edges incoming to $v$ and
$e'_1,\dots,e'_m$ are the edges outgoing from $v$.
\end{enumerate}
\end{defn}

Now let us go up a dimension and define `spin foams'.  A 2-dimensional
oriented complex has a finite set of vertices $V$, a finite set of edges
$E$, and finite sets of $n$-sided 2-cells or {\it faces} $F_n$  for each
$n \ge 3$, with only finitely many $F_n$ being nonempty.  As in a
1-dimensional oriented complex, the orientations of the edges give maps
\[         s,t \maps E \to V \]
assigning to each edge its source and target.  In addition, the
orientation on any 2-cell $f \in F_n$ puts a cyclic ordering on its
faces and vertices.  Suppose we arbitrarily choose a `first' vertex
for each 2-cell $f$ of our complex.  Then we may number all its vertices
and edges from $1$ to $n$.
It is convenient to think of these numbers as lying in $\Z_n$.  

We thus obtain maps
\[        e_i \maps F_n \to E, \quad v_i \maps F_n \to V 
\qquad \qquad i \in \Z_n. \]
Note that for each $f \in F_n$ either 
\be       s(e_i(f)) = v_i(f) \;{\rm and}\; t(e_i(f)) = v_{i+1}(f) 
\label{incoming} \ee
or 
\be       t(e_i(f)) = v_i(f) \;{\rm and}\;\; s(e_i(f)) = v_{i+1}(f).
\label{outgoing} \ee
If (\ref{incoming}) holds, we say $f$ is {\it incoming} to $e$,
while if (\ref{outgoing}) holds, we say $f$ is {\it outgoing} from
$e$.  In other words, $f$ is incoming to $e$ if the orientation
of $e$ agrees with the orientation it inherits from $f$, while it is
outgoing if these orientations do not agree.

First we define a special class of spin foams:
\begin{defn}\et 
A {\it closed spin foam} $F$ is a triple $(\kappa,\rho,\iota)$
consisting of:
\begin{enumerate}
\item a 2-dimensional oriented complex $\kappa$, 
\item a labeling $\rho$ of each face $f$ of $\kappa$ by an 
irreducible representation $\rho_f$ of $G$,
\item a labeling $\iota$ of each edge $e$ of $\kappa$ by an intertwiner
\[       \iota_e \maps \rho_{f_1} \tensor \cdots \tensor \rho_{f_n} 
\to \rho_{f'_1} \tensor \cdots \tensor \rho_{f'_m}  \]
where $f_1,\dots,f_n$ are the faces incoming to $e$ and
$f'_1,\dots,f'_m$ are the faces outgoing from $e$.  
\end{enumerate}
\end{defn}

Next we turn to general spin foams.  In general, a spin foam $F \maps
\Psi \to \Psi'$ will go from a spin network $\Psi$ to a spin network
$\Psi'$.
It has `free edges', the edges of the spin networks $\Psi$ and
$\Psi'$, which are not labeled by intertwiners.  It also has edges
ending at the spin network vertices, and the intertwiners labeling
these edges must match those labeling the spin network vertices.  A
closed spin foam is just a spin foam of the form $F \maps \emptyset \to
\emptyset$, where $\emptyset$ is the {\it empty spin network}: the spin
network with no vertices and no edges.  

To make this more precise, suppose $\gamma$ is a 1-dimensional oriented
complex and $\kappa$ is a 2-dimensional oriented complex.  Note that the
product $\gamma \times [0,1]$ becomes a 2-dimensional oriented complex
in a natural way.  We say $\gamma$ {\it borders} $\kappa$ if there is a
one-to-one affine map $c \maps |\gamma| \times [0,1] \to |\kappa|$ mapping
each cell of $\gamma \times [0,1]$ onto a unique cell of $\kappa$ in an
orientation-preserving way, such that $c$ maps $\gamma \times [0,1)$
onto an open subset of $|\kappa|$.   Note that in this case, $c$ lets us
regard each $j$-cell of $\gamma$ as the face of a unique $(j+1)$-cell
of $\kappa$.  
Each vertex $v$ of $\gamma$ is the source or target of a unique edge of
$\kappa$, which we denote by $\tilde v$, and each edge $e$ of $\gamma$ is
the edge of a unique face of $\kappa$, which we denote by $\tilde e$.

It is easiest to first define spin foams $F \maps \emptyset 
\to \Psi$ and then deal with the general case:

\begin{defn}\et
Suppose that $\Psi = (\gamma,\rho,\iota)$ is a spin network.  
A {\it spin foam} $F \maps \emptyset \to \Psi$ 
is a triple $(\kappa,\tilde \rho,\tilde \iota)$ consisting of:
\begin{enumerate}
\item a 2-dimensional oriented complex $\kappa$ such that
$\gamma$ borders $\kappa$, 
\item a labeling $\tilde \rho$ of each face $f$ of $\kappa$ by 
an irreducible representation $\tilde \rho_f$ of $G$,
\item a labeling $\tilde \iota$ of each edge $e$ of $\kappa$ not 
lying in $\gamma$ by an intertwiner
\[      \tilde\iota_e \maps \rho_{f_1} \tensor \cdots \tensor \rho_{f_n}
\to \rho_{f'_1} \tensor \cdots \tensor \rho_{f'_m}  \]
where $f_1,\dots,f_n$ are the faces incoming to $e$ and
$f'_1,\dots,f'_m$ are the faces outgoing from $e$.
\end{enumerate}
such that the following hold:
\begin{enumerate}
\item For any edge $e$ of $\gamma$, $\tilde \rho_{\tilde e} = \rho_e$
if $\tilde e$ is incoming to $e$, while $\tilde \rho_{\tilde e} = 
(\rho_e)^\ast$ if $\tilde e$ is outgoing to $e$.
\item For any vertex $v$ of $\gamma$, $\tilde \iota_{\tilde e}$ 
equals $\iota_e$ after appropriate dualizations.  
\end{enumerate}
\end{defn}

To define general spin foams, we need the notions of `dual' and `tensor
product' for spin networks.  Suppose $\Psi = (\gamma,\rho,\iota)$ is a
spin network.   Then the {\it dual} of $\Psi$ is the spin network
$\Psi^\ast$ with the same underlying 1-dimensional oriented complex 
$\gamma$, but with each edge $e$ labeled by the representation
$\rho_e^\ast$, and with each vertex $v$ labeled by the appropriately
dualized form of the the intertwining operator $\iota_v$.   

Suppose that $\Psi = (\gamma,\rho,\iota)$ and $\Psi' = (\gamma', \rho',
\iota')$ are disjoint spin networks in the same space $\R^n$.  Then the
{\it tensor product} $\Psi \tensor \Psi'$ is defined to be the spin
network whose underlying 1-dimensional oriented complex is the disjoint
union of $\gamma$ and $\gamma'$, with edges and vertices labeled by
representations and intertwiners using $\rho,\rho'$ and $\iota,\iota'$. 
 
\begin{defn}\et
Given disjoint spin networks $\Psi$ and $\Psi'$ in $\R^n$
a {\it spin foam} $F \maps \Psi \to \Psi'$ is defined
to be a spin foam $F \maps \emptyset \to  \Psi^\ast \tensor \Psi'$.
\end{defn}
 
Our notation here is meant to suggest that there is a category with
spin networks as objects and spin foams as morphisms.  In other words, we
want to be able to compose spin foams $F \maps \Psi \to \Psi'$ and $F'
\maps \Psi' \to \Psi''$ and obtain a spin foam $FF' \maps \Psi \to
\Psi''$.   We want this composition to be associative, and for each spin
network $\Psi$ we want a spin foam $1_\Psi \maps \Psi \to \Psi$ serving
as a left and right unit for composition.   

To get this to work, one must deal with some technicalities similar to
those that arise when one defines the category of cobordisms in
topological quantum field theory \cite{Sawin}.  Say one has spin foams
$F \maps \Psi \to \Psi'$ and $F \maps \Psi' \to  \Psi''$ with $F =
(\kappa,\rho,\iota)$ and $F' = (\kappa',\rho',\iota')$.  One would like
to define $FF'$ to be the result of gluing $\kappa$ to $\kappa'$ and
labeling the resulting complex with the representations coming from
$\rho,\rho'$ and the intertwiners coming from $\iota,\iota'$.   However,
there is no guarantee that $\kappa$ and $\kappa'$ are disjoint, or that
they lie in the same space $\R^n$.   We can arbitrarily choose disjoint
copies of them lying in the same space, but then composition satisfies
associativity and the left and right unit laws only up to a certain
equivalence relation on spin foams.  To get a category, one must
therefore take the morphisms to be certain equivalence classes of spin
foams.  We give the details in Appendix B.  

We conclude this section with a few remarks:

1) There is a close analogy between the category of spin foams and the
category of cobordisms used in topological quantum field theory.  This
analogy is heightened if we use the the trivial group as our group $G$. 
Then a spin network is just a 1-dimensional oriented complex, which we
can think of as a model of `space', and a spin foam is a 2-dimensional
oriented complex which we can think of as a `spacetime'.  Using a
nontrivial group amounts to equipping these complexes with certain extra
labelings, which we can think of as `fields'.  

2) In addition to the `closed' spin networks as defined above, we could
define more general `open' spin networks with free vertices labeled
not by an intertwining operator but by a vector in the representation
labeling the incident edge.  As noted in the Introduction, open spin
networks are closely related to Feynman diagrams.  They also arise
naturally in the study of gauge theory on a manifold with boundary, where
they are embedded in the manifold with their free vertices on
the boundary.  An example of this is the spin network description of a
quantum black hole, where the event horizon plays the role of a boundary
\cite{ABCK}.  Another example is the spin network description of the
asymptotically flat sector of quantum gravity, where the `boundary' is
the sphere at spacelike infinity \cite{T3}.  To study the dynamics of
such situations, it might be useful to introduce spin foams going
between open spin networks.  

3) We could also define spin networks and spin foams in a more general,
purely combinatorial way.  Abstracting from the notion of a
1-dimensional oriented complex, one may define a {\it graph} to be a
pair of finite sets $E,V$ together with maps $s,t \maps E \to V$.  Every
1-dimensional oriented complex has an underlying graph, and the complex
is determined by its graph, at least up to piecewise linear
homeomorphism (as explained by Rourke and Sanderson \cite{RoSa}).
Conversely, a graph is the underlying graph of a 1-dimensional oriented
complex if and only if no vertex is both the source and target of the
same edge.  One can easily generalize the definition of a spin network
$\Psi = (\gamma,\rho,\iota)$ by allowing $\gamma$ to be an arbitrary
graph.

Similarly, abstracting from the notion of a 2-dimensional oriented
complex, one may define a {\it 2-graph} to consist of finite sets $V,E,$
and $F_n$ for $n \ge 3$, with only finitely many $F_n$ nonempty,
together with maps $s,t \maps E \to V$ and $e_i \maps F_n \to E$, $v_i
\maps F_n \to V$ for $i \in \Z_n$, such that for any $f \in F_n$
either (\ref{incoming}) or (\ref{outgoing}) holds.  A 2-dimensional
oriented complex determines a 2-graph if we arbitrarily choose a `first'
vertex for each face.  We can reconstruct the complex from this 2-graph,
at least up to piecewise linear homeomorphism \cite{RoSa}.  Presumably
there are purely combinatorial conditions on a 2-graph implying that it
comes from a 2-dimensional oriented complex.  It is certainly necessary
that no vertex should be the source and target of the same edge, and
that no edge should be both the $i$th and $j$th edge of the same face
for $i \ne j$.  However, the author does not know necessary and
sufficient conditions.  In any event, one can generalize the definition
of a spin foam $F = (\kappa,\rho,\iota)$ by allowing $\kappa$ to be an
arbitrary 2-graph.

4) One could also consider higher-dimensional analogs of spin networks
and spin foams.  For $BF$ theory and the connection formulation of
quantum gravity in three and four dimensions, the basic fields can all be
thought of as vector-bundle-valued differential forms of degree $\le 2$.
As we shall see, this suggests that the quantization of these theories
should involve only 2-dimensional complexes.  Work on `$p$-branes' and
the like suggests that some field theories involving forms of higher
degree are classical limits of theories involving higher-dimensional
complexes.

\section{Spin foams in lattice gauge theory} \label{lattice.gauge}

In certain lattice gauge theories, spin networks describe states, while
spin foams describe `histories': the path integral can be computed as a
sum over spin foams.  In this context we work, not with the abstract
spin networks of the previous section, but with spin networks embedded
in a manifold representing space.  Similarly, we work with spin foams
embedded in a manifold representing spacetime.  Throughout the rest of
the paper we assume these manifolds are compact, oriented, and equipped
with a fixed triangulation.  The triangulation specifies another
decomposition of the manifold into cells called the `dual complex'.
There is a one-to-one correspondence between $k$-simplices in the
triangulation of an $n$-manifold and $(n-k)$-cells in the dual complex,
each $k$-simplex intersecting its corresponding dual $(n-k)$-cell in a
single point.  Our spin networks and spin foams live in the appropriate
dual complexes, as sketched in the Introduction.  We need to work with
oriented complexes, so we orient each cell of the dual complex in an
arbitrary fashion.

We begin by recalling how spin networks describe states in lattice gauge
theory.  We fix a compact connected Lie group $G$ as our gauge
group, and suppose $S$ is an $(n-1)$-manifold representing `space'.  We
assume $S$ is equipped with a triangulation $\Delta$ and a principal
$G$-bundle $P \to S$.  Also, we choose a trivialization of $P$ over every
0-cell of the dual complex $\Delta^\ast$.

The `$k$-skeleton' of a complex is the subcomplex formed by all cells of
dimension less than or equal to $k$.  The 1-skeleton of $\Delta^\ast$ is
a graph, and we can set up gauge theory on this graph in the usual way
\cite{B1.5}.   We represent parallel transport along each edge of
$\Delta^\ast$ as an element of $G$, so we define the space of {\it
connections} $\A_S$ by 
\[      \A_S = \prod_{e \in \Delta_1^\ast}   G  \] 
where $\Delta_k^\ast$ denotes the set of $k$-cells in $\Delta^\ast$.  
Similarly, we represent a gauge transform at each vertex of
$\Delta^\ast$ as an element of $G$, so we define the group of {\it gauge
transformations} $\G_S$ by  
\[   \G_S = \prod_{v \in \Delta_0^\ast} G  .\] 
The group $\G_S$ acts on $\A_S$, and the quotient space $\A_S/\G_S$ is
the space of connections modulo gauge transformations in this
setting.  The space $\A_S$ has a measure on it given by a product of
copies of Haar measure, and this measure pushes forward to one on
$\A_S/\G_S$.  Using this we define the Hilbert space $L^2(\A_S/\G_S)$.  

Suppose that $\Psi = (\gamma,\rho,\iota)$ is a {\it spin network in $S$},
that is, one for which $\gamma$ is the 1-skeleton of $\Delta^\ast$. 
Then by a now familiar argument \cite{B3}, $\Psi$ defines a state in
$L^2(\A_S/\G_S)$, which we also call $\Psi$.  Moreover, such {\it spin
network states} span $L^2(\A_S/\G_S)$.  In fact, we obtain an
orthonormal basis of states as $\rho$ ranges over all labelings of
the edges of $\Delta^\ast$ by irreducible representations of $G$ and
$\iota$ ranges over all labelings of the vertices by intertwiners
chosen from some orthonormal basis.  

Next we turn to spin foams.  For this, suppose $S$ is a submanifold of
some $n$-manifold $M$ representing spacetime.   We assume that $P \to M$
is a principal $G$-bundle over $M$ that restricts to the already given
bundle over $S$, and that $\Theta$ is a triangulation of $M$ that
restricts to the already given triangulation of $S$.   We wish to see
how a closed spin foam in $M$ determines a spin network in $S$.  

Suppose that $F = (\kappa,\tilde \rho,\tilde \iota)$ is a {\it closed spin
foam in $M$}, that is, one for which $\kappa$ is the 2-skeleton of
$\Theta^\ast$.  Note that every $k$-cell $X$ of $\Delta^\ast$ is
contained in a unique $(k+1)$-cell $\tilde X$ of $\Theta^\ast$.  Since
$S$ and $M$ are oriented, the normal bundle of $S$ acquires an
orientation.   Similarly, since $X$ and $\tilde X$ are oriented, the
normal bundle of the interior of $X$ in $\tilde X$ acquires an
orientation.   But this latter bundle can be identified with the
restriction of the normal bundle of $S$ to the interior of $X$, so we
have two different orientations on the same bundle.   We say that
$\tilde X$ is {\it incoming} to $X$ if these orientations agree, and
{\it outgoing} to $X$ if they do not.  We thus obtain a spin network in
$S$ as follows:

\begin{prop}\et If $F = (\kappa,\tilde \rho,\tilde \iota)$ is a closed
spin foam in $M$, there exists a unique spin network $F|_S =
(\gamma,\rho,\iota)$ in $S$ such that:

\begin{enumerate}
\item $\gamma$ is the 1-skeleton of $\Delta^\ast$,
\item for any edge $e$ of $\gamma$, $\tilde \rho_{\tilde e} =
\rho_e$ if $\tilde e$ is incoming to $e$, while $\tilde \rho_{\tilde e}
= (\rho_e)^\ast$ if $\tilde e$ is outgoing to $e$. 
\item If $v$ is a vertex of $\gamma$, then $\tilde \iota_{\tilde e}$ 
equals $\iota_e$ after appropriate dualizations.   
\end{enumerate}

\end{prop}

\noindent The proof is trivial but the physical idea is important: a
history on spacetime determines a state on any submanifold
corresponding to space at a given time. 

A simpler version of this idea applies when we have a spin foam in a
cobordism $M \maps S \to S'$.   Here we assume that $M$ has a boundary
which has been identified with the disjoint union of $S$ and $S'$.   We
assume there is a principal $G$-bundle $P \to M$ and a triangulation
$\Theta$ of $M$ which restricts to triangulations $\Delta$ and
$\Delta'$ of $S$ and $S'$, respectively.  The concept of dual complex
is a bit subtler for manifolds with boundary: the dual complex
$\Theta^\ast$ has one $k$-cell for each $(n-k)$-cell of
$\Theta$, but also includes all the cells of $\Delta^\ast$ and
$\Delta'^\ast$.   We fix a trivialization of $P$ at every vertex of
$\Theta^\ast$ on the boundary of $M$.   

Note that the 1-skeleton of $\Delta^\ast \cup {\Delta'}^\ast$ borders
the 2-skeleton of $\Theta^\ast$, in the sense defined in the previous
section.  We define a {\it spin foam in $M$} to be a spin foam $F
\maps \Psi \to \Psi'$ such that the underlying complex of $F$ is the
2-skeleton of $\Theta$, $\Psi$ is a spin network in $S$, and $\Psi'$ is
a spin network in $S'$.  Note that this reduces to the previous notion
of `spin foam in $M$' when the boundary of $M$ is empty.

Any spin foam $F \maps \Psi \to \Psi'$ in $M$ determines an operator
from $L^2(\A_S/\G_S)$ to $L^2(\A_{S'}/\G_{S'})$, which we
also denote by $F$, such that
\[          \langle \Phi', F\Phi \rangle = \langle \Phi',\Psi' \rangle
\langle \Psi,\Phi\rangle \]
for any states $\Phi,\Phi'$.  It may seem odd to call this operator
`$F$', since it depends only on $\Psi$ and $\Psi'$, not any other
details of the spin foam.  The point is that the spin foam $F$
represents a history going from the initial state $\Psi$ to the final
state $\Psi'$, and the corresponding operator does not depend on the
behavior of this history at `intermediate times', that is, in the
interior of $M$.  

We call the operator $F$ a {\it spin foam operator}.  Just as the space
of states $L^2(\A_S/\G_S)$ is spanned by spin network states, every
operator from $L^2(\A_S/\G_S)$ to $L^2(\A_{S'}/\G_{S'})$ will be a
linear combination of spin foam operators if there is a spin foam $F
\maps \Psi \to \Psi'$ for every pair of spin networks $\Psi$ and
$\Psi'$.  In a wide variety of lattice gauge theories, the time
evolution operator can be computed as a sum of spin foam operators,
weighted by amplitudes computed as a product of vertex, edge, and face
amplitudes.  For more on how this works in Yang-Mills theory and $BF$
theory, see the paper by Reisenberger \cite{R1} and the references
therein.  $BF$ theory with gauge group $\SU(2)$ is 3-dimensional
Euclidean quantum gravity, and the lattice formulation of this theory is
just the Ponzano-Regge model \cite{PR}.  When the cosmological constant
$\Lambda$ is nonzero, one obtains instead the Turaev-Viro model
\cite{TV}, where the quantum group $\SU_q(2)$ takes the place of the
gauge group $\SU(2)$, with the deformation parameter $q$ being a
function of $\Lambda$.  In the rest of this paper we concentrate on a
more complicated example, namely 4-dimensional Euclidean quantum
gravity.

\section{Spin networks as quantum 3-geometries} \label{3.geom}

In this section and the next, our goal is to understand how a linear
combination of $\SU(2)$ spin foams in a triangulated 4-manifold can
endow it with a `quantum 4-geometry'.  The basic picture is simple: spin
foam faces give area to the triangles they intersect, while spin foam
edges give 3-volume to the tetrahedra they intersect.  We also suspect
that spin foam vertices give 4-volume to the 4-simplices in which they
lie.  However, the 3-dimensional aspects of this picture are much better
understood than the 4-dimensional ones.  Thus, before turning to the
question of how spin foams represent quantum 4-geometries, let us recall
how spin networks represent quantum 3-geometries.  Little in this
section is completely new: the main task is to put existing ideas
together in a coherent picture.

Let $S$ be a compact oriented 3-manifold equipped with a fixed triangulation
$\Delta$ and a principal $\SU(2)$ bundle $P \to S$.   
Starting from this bundle we can construct the space $L^2(\A_S/\G_S)$ as
in the previous section.  Let $\Delta_k$ stand for the set of
$k$-simplices in $\Delta$.  Note that the dual complex $\Delta^\ast$ has
one 4-valent vertex for every tetrahedron in $\Delta$ and one edge for
every triangle in $\Delta$.  Given a tetrahedron $T \in \Delta_3$, let
$\partial_0 T, \dots , \partial_3 T$ denote its faces, or equivalently,
the corresponding edges of the dual complex.  From the results of the
previous section we have:
\[ L^2(\A_S/\G_S) \iso \bigoplus_j \bigotimes_{T \in \Delta_3}  
\Inv(j_{\partial_0 T} \otimes j_{\partial_1 T} \otimes 
j_{\partial_2 T} \otimes j_{\partial_3 T})  \] 
where `$\Inv$' denotes the subspace of vectors transforming
in the trivial representation, and 
$j$ ranges over all labelings of the triangles of $\Delta$ by
spins.  Note that to obtain this isomorphism we need to use the fact
that every representation of $\SU(2)$ is equivalent to its dual.

The work of Barbieri \cite{Barb} clarifies the geometrical significance
of this description of $L^2(\A_S/\G_S)$: the space $\Inv(j_0 \tensor j_1
\tensor j_2 \tensor j_3)$ describes the states of a `quantum
tetrahedron' whose $i$th face has area ${1\over 2}\sqrt{j_i(j_i + 1)}$. 
We review this work in what follows, but first, note an important
consequence: the above equation describes $L^2(\A_S/\G_S)$ as the {\it
Hilbert space of quantum 3-geometries} of the simplicial complex
$\Delta$.  Labeling the triangles of $\Delta$ by spins fixes the areas
of the triangles and determines a space of states consistent with these
areas for each tetrahedron in $\Delta$.   To obtain $L^2(\A_S/\G_S)$ we
take the tensor product of all these spaces and then the direct sum over
all labelings. 

To understand Barbieri's work on the quantum tetrahedron, start with a
tetrahedron $T$ having vertices $0,1,2,3$.  We can think of a
tetrahedron in $\R^3$ with one vertex at the origin as an affine map
from $T$ to $\R^3$ sending the vertex $0$ to the origin.   Such a map is
given by fixing vectors $e_1,e_2,e_3 \in \R^3$ corresponding to the
edges $01$, $02$, and $03$.  
This data is precisely a `cotriad', a linear map from the tangent space
of the vertex $0$ of $T$ to $\R^3$.  In the Palatini approach to general
relativity one works with a similar object, namely the `cotriad field',
which can be thought of locally as an $\R^3$-valued 1-form $e$.  But a
crucial idea in modern canonical quantum gravity is to work not with $e$
but with the $\Lambda^2 \R^3$-valued 2-form $e \we e$, often called the
`densitized cotriad field'.  Using this rather than the cotriad field is
the reason why areas rather than lengths are the basic geometrical
observables in the spin network approach to quantum gravity.  This
suggests an alternate description of the geometry and orientation of a
tetrahedron in terms of faces rather than edges.

In this alternate description, we work not with the vectors $e_i$ but
with the bivectors $e_i \we e_j$.  (A bivector in $n$ dimensions is
simply an element of $\Lambda^2 \R^n$; in the presence of a metric a
bivector in three dimensions can be identified with an `axial vector', 
which in turn can be identified with a vector in the presence of an
orientation.)   We denote these bivectors by: 
\[          E_1 = e_2 \we e_3 , \qquad
            E_2 = e_3 \we e_1, \qquad
            E_3 = e_1 \we e_2 . \] 
Not every triple of bivectors $E_i$ comes from a triple of vectors $e_i$
this way.   It is necessary and sufficient that either the 
$E_i$ are linearly independent and satisfy
\[        \epsilon_{IJK} E_1^I E_2^J E_3^K > 0, \]
or that they are all multiples of a fixed bivector.
The first case is the generic one; it occurs whenever the vectors
$e_i$ are linearly independent.  In this case, the $E_i$ determine
the $e_i$ up to a parity transformation $e_i \mapsto -e_i$.  The second
case occurs when the vectors $e_i$ lie in a plane; in this case the
$E_i$ determine the $e_i$ up to an area-preserving (but possibly
orientation-reversing) linear transformation of this plane.  

In short, the map from triples $e_i$ to triples $E_i$ is neither
one-to-one nor onto.   Barbieri has suggested an interesting way
around this, at least in the generic case.   This trick involves fixing
an orientation on $\R^3$.  If the linearly independent triple $e_i$ is
right-handed, we define the $E_i$ as above, but if the triple $e_i$ is
left-handed, we define the $E_i$ with a minus sign:
\[          E_1 = -e_2 \we e_3 , \qquad 
            E_2 = -e_3 \we e_1, \qquad
            E_3 = -e_1 \we e_2 , \] 
This establishes a one-to-one and onto map from linearly independent
triples $e_i$ to linearly independent triples $E_i$.  While somewhat
problematic, we tentatively adopt this strategy in what follows, since
it allows us the convenience of working with arbitrary triples $E_i$
if we simply accept the fact that a linearly dependent triple $E_i$
does not describe a unique tetrahedron.

Actually, rather than working directly with a triple of bivectors, it is 
useful to introduce a fourth:   
\[          E_0 = -E_1 - E_2 - E_3 . \]
Then, if we use a metric and a orientation to think of the $E_i$ as
vectors, each $E_i$ is normal to the $i$th face of the tetrahedron (the
face missing the $i$th vertex), with magnitude given by the area of that
face.   We can obtain either inwards or outwards normals this way.
In this description, the constraint $E_0 + E_1 + E_2 + E_3 = 0$ is a
consequence of the fact that the integral of the unit normal to a
2-sphere piecewise linearly embedded in $\R^3$ must vanish.  We call it
the {\it closure constraint}, since it says that the faces can close to
form a tetrahedron.  

To `quantize' the tetrahedron, Barbieri first quantizes the bivectors
$E_i$ and then imposes the closure constraint.  The notion of `quantized
bivectors' is implicit in Penrose's ground-breaking work on spin
networks and `quantized directions' \cite{Moussouris,Penrose}, but
quantized bivectors really date back to early quantum mechanics, where
they were introduced to describe angular momentum.  Classically, the
angular momentum is a bivector.  Quantum mechanically, the components of
the angular momentum no longer commute, but instead satisfy
\[       [J^1,J^2] = iJ^3,\qquad 
         [J^2,J^3] = iJ^1,\qquad 
         [J^3,J^1] = iJ^2  .\] 
The uncertainty principle thus prevents us from simultaneously measuring
all 3 components with arbitrary accuracy.  Moreover, each component 
takes on a discrete spectrum of values.   

This is all very elementary, but a more sophisticated viewpoint is also
helpful.  We can use a metric on $\R^3$ to identify $\Lambda^2 \R^3$
with $\so(3)^\ast$, which is the phase space of a classical spinning
point particle.  Any element $f \in \so(3)$ gives a linear function on
$\so(3)^\ast$, essentially the angular momentum about a certain axis.
The space $\so(3)^\ast$ is naturally equipped with the `Kirillov-Kostant
Poisson structure', in which the Poisson brackets of any two such linear
functions is given by
\[ \{f,g\} = [f,g] .\] 
Being odd-dimensional, $\so(3)^\ast$ is not a symplectic manifold, but
it is foliated by symplectic leaves, namely the concentric spheres about
the origin, which are just the orbits of the action of $\SO(3)$ --- the
so-called `coadjoint orbits'.  Following Kirillov and Kostant, we can
apply geometric quantization to this phase space \cite{GS}.  Only the
`integral' coadjoint orbits contribute, that is, those for which
symplectic 2-form divided by $2\pi$ defines an integral cohomology
class.  These are the spheres of radius $j = 0, {1\over 2}, 1, \dots\,$.
Over each such sphere there is a complex line bundle having the
symplectic 2-form as its curvature, and if we think of this sphere as
the Riemann sphere, the space of holomorphic sections of this line
bundle is naturally isomorphic to the spin-$j$ representation of
$\so(3)$.  The direct sum of all these spaces is the Hilbert space of a
quantum spinning point particle, or in other words, the {\it Hilbert
space of a quantum bivector in three dimensions}.  We denote this space
by 
\[ {\cal H} = \bigoplus_j j \] 
where we denote the spin-$j$ representation of $\so(3)$ simply by $j$.
This space is a representation of $\so(3)$, so there are operators
$J^1,J^2,J^3$ on it satisfying the above commutation relations.  Note
that this space is not a represention of $\SO(3)$, but only of its
universal cover, $\SU(2)$.  In this way quantum bivectors are naturally
related to $\SU(2)$ gauge theory.

To construct the Hilbert space of the quantum tetrahedron, Barbieri
starts with ${\cal H}^{\tensor 4}$, the tensor product of four copies 
of the Hilbert space for a quantum bivector, one for
each face of tetrahedron.  On this space we have operators 
\ban \hat E_0^I = J^I \tensor 1 \tensor 1 \tensor 1  \\
     \hat E_1^I = 1 \tensor J^I \tensor 1 \tensor 1  \\
     \hat E_2^I = 1 \tensor 1 \tensor J^I \tensor 1  \\
     \hat E_3^I = 1 \tensor 1 \tensor 1 \tensor J^I  \ean
where $I = 1,2,3$.   The {\sl Hilbert space of the quantum tetrahedron}
is then defined as the subspace
\[   {\cal T} =  \{ \psi \in {\cal H}^{\tensor 4}\; \colon \;
(\hat E_0 + \hat E_1 + \hat E_2 + \hat E_3)\psi = 0 \}  \]
on which the quantized version of the closure constraint holds.   Since
${\cal H}^{\tensor 4}$ is naturally a representation of $\so(3)$ and the
components of the closure constraint generate the $\so(3)$ action, it
follows that
\[     {\cal T} \iso \bigoplus_{j_0,j_1,j_2,j_3}
\Inv(j_0 \tensor j_1 \tensor j_2 \tensor j_3)  .\]

Classically the area of the $i$th face of the tetrahedron is given by
\[          A_i = {1\over 2} (E_i \cdot E_i)^{1/2}. \]
Quantizing these expressions, we define {\it area operators} on ${\cal T}$ 
by  
\[     \hat A_i = {1\over 2} (\hat E_i \cdot \hat E_i)^{1/2} \]
for $i = 0,1,2,3$.   These operators  commute and are simultaneously
diagonalized on the subspaces $\Inv(j_0 \tensor \cdots \tensor j_3)$;
the eigenvalue of $\hat A_i$ on this subspace is ${1\over
2}\sqrt{j_i(j_i + 1)}$.   Similarly, the volume $V$ of a tetrahedron is
given classically in terms of the bivectors $E_i$ by 
\[         V = {1\over 6}|\epsilon_{IJK} E_1^I E_2^J E_3^K|^{1/2} ,\]
so we define the {\it volume operator} on ${\cal T}$ by
\[ \hat V = 
{1\over 6} |\epsilon_{IJK} \hat E_1^I \hat E_2^J \hat E_3^K|^{1/2} .\]
More precisely, this operator is first defined on ${\cal H}^{\tensor 4}$,
but since it commutes with the action of $\so(3)$ on this space it
preserves the subspace $\cal T$.  (In the above formulas, we
have ignored factors involving $c, \hbar, 8\pi G$, and the Immirzi 
parameter \cite{Im}.)

It follows that on the space
\[   L^2(\A_S/\G_S) \iso \bigoplus_j  \bigotimes_{T \in \Delta_3} 
\Inv(j_{\partial_0 T} \otimes j_{\partial_1 T} 
\otimes j_{\partial_2 T} \otimes j_{\partial_3 T})  \]
we can define an area operator $\hat A_f$ for each triangle $f \in
\Delta_2$ and a volume operator $\hat V_T$ for each tetrahedron $T \in
\Delta_3$.   All these operators commute, and on the subspace of
$L^2(\A_S/\G_S)$ corresponding to a given labeling $j$, each area operator
$\hat A_f$ has eigenvalue ${1\over 2}\sqrt{j_f(j_f + 1)}$.

The reader may find it curious how imposing the closure constraint
automatically restricts attention to rotation-invariant aspects of the
geometry of the quantum tetrahedron.  In fact,  the constraint $E_0 +
\cdots + E_3 = 0$ is a kind of simplicial version of the Gauss
constraint $d_A E = 0$.  Since the Gauss 
constraint generates $\SU(2)$ gauge transformations, it should not be 
surprising that the closure constraint generates rotations of the 
tetrahedron.    Imposing the closure constraint for a given 
tetrahedron of $\Delta$ has the same effect as imposing gauge-invariance 
at the spin network vertex centered at that tetrahedron.  

Making this precise requires a bit of care.  We cannot obtain
$L^2(\A_S/\G_S)$ simply by starting with a tensor product of copies of
$\cal H$, one for each triangle in $\Delta$, and then imposing a closure
constraint for each tetrahedron.  This gives too small a space of
states.  Instead, we must first `explode' the triangulation $\Delta$
into a disjoint union of tetrahedra.  The resulting complex has two
triangles for every triangle in $\Delta$.  If we take a tensor product
of one copy of $\cal H$ for each triangle in this `exploded' complex, we
obtain the Hilbert space
\[     H(S) = \bigotimes_{T \in \Delta_3} {\cal H}^{\otimes 4} .\] 
On this space we have operators $\hat E_i^I(T)$ 
generating $\so(3)$ actions on each copy of $\cal H$, one for
each face $i = 0,1,2,3$ of  each tetrahedron $T$ in $\Delta$.  These
give area operators  
\[   \hat A_i(T) = {1\over 2} (\hat E_i(T) \cdot \hat E_i(T))^{1/2}. \]

Starting from this big Hilbert space we can impose constraints as
follows to get down to $L^2(\A_S/\G_S)$.   First, since each
triangle in $\Delta$ corresponds to a pair of triangles in the
`exploded' complex, we can impose constraints saying that both triangles
in each such pair have the same area.  In other words, we can take 
the subspace of states $\psi \in H(S)$ for which
\[      A_i(T)\psi = A_{i'}(T')\psi  \]
whenever $T,T'$ are tetrahedra in $\Delta$ with $\partial_i T 
= \partial_{i'} T'$.  We obtain the subspace
\ban \bigoplus_j \bigotimes_{T \in \Delta_3}  
j_{\partial_0 T} \otimes j_{\partial_1 T} \otimes 
j_{\partial_2 T} \otimes j_{\partial_3 T} &\iso&  
\bigoplus_j \bigotimes_{f \in \Delta_2} j_f \tensor j_f \\
&\iso& \bigotimes_{f \in \Delta_2} L^2(SU(2))  \\
&\iso& L^2(\A_S).  \ean
where $j$ ranges over all labelings of triangles in $\Delta$ by spins.  

Next, we can impose a copy of the closure constraint for each tetrahedron,
picking out states $\psi$ for which   
\[  (\hat E_0(T) + \hat E_1(T) + \hat E_2(T) + \hat E_3(T)) \psi  = 0 \]
for all $T$ in $\Delta$.  This gives the subspace 
\[ \bigoplus_j \bigotimes_{T \in \Delta_3}  
\Inv(j_{\partial_0 T} \otimes j_{\partial_1 T} \otimes 
j_{\partial_2 T} \otimes j_{\partial_3 T}) \iso L^2(\A_S/\G_S) \]
as desired.  

We can also impose these constraints in the other order.  If starting
from $H(S)$ we impose a copy of the closure constraint for each
tetrahedron, we obtain the subspace  $\bigotimes_{T \in \Delta_3} {\cal
T}$.   Then, imposing equal-area constraints for pairs of triangles  in
the `exploded' complex, we obtain $L^2(\A_S/\G_S)$.  In short, the big
space $H(S)$ contains both $L^2(\A_S)$ and  $\bigotimes_{T \in \Delta_3}
{\cal T}$, and the intersection of these two spaces is $L^2(\A_S/\G_S)$.

We conclude with a few remarks:

1)  It is interesting to compare the area and volume operators 
discussed above with others appearing in the quantum gravity
literature.  Most previous work has been done (at least implicitly) in
the real-analytic rather than the piecewise-linear context.   The area
and volume operators first introduced by Rovelli and Smolin \cite{RS2}
differ from those later studied by Ashtekar, Lewandowski and
collaborators \cite{AL2.5,AL3,AL4,ALMMT,L}.  However, these differences do not
arise in situations involving spin networks with generic 4-valent
vertices (i.e., those for which no triple of incident edges have
linearly dependent tangent vectors).  In such situations, their formulas
agree with the formulas above up to constant factors.  

Loll works in the piecewise-linear context, but she considers 6-valent
spin networks dual to a cubical lattice \cite{Loll,Loll2,Loll3,Loll4}.
If one sets to zero two of the spins labeling the six edges incident to a
vertex, her area and volume operators match those above up to constant
factors, except for one difference: she does not include an absolute
value in the definition of the volume operator.  Instead, she notes
that the nonzero eigenvalues of the operator $\epsilon_{IJK} \hat
E_1^I \hat E_2^J \hat E_3^K$ come in pairs of the same magnitude but
opposite sign, and proposes eliminating the subspace on which this
operator is negative.  As described above, Barbieri's work suggests
another way out of this problem: changing the definition of $E_i$ in
terms of the cotriad field $e_i$, so that the above quantity is
negative when the cotriad is left-handed.  Of course, this strategy
cannot be blithely adopted without more study.  It would also be
worthwhile to better understand the constant factors by which the area
and volume operators used here differ from those appearing elsewhere.

2)  An alternate approach to the quantum tetrahedron involves
imposing the closure constraint at the classical level.  Here we start
with the product of four copies of $\so(3)^\ast$, equipped with its product
Poisson structure.  This space is foliated by 8-dimensional symplectic
leaves of the form $S^2 \times S^2 \times S^2 \times S^2$, where the
radius $j_i$  of the $i$th sphere is the area of the $i$th face of the
tetrahedron.   The symplectic structure on a leaf is integral precisely
when each $j_i$ equals $0,{1\over 2},1,\dots\,$.  To impose the closure
constraint we apply symplectic reduction to each leaf \cite{GS}, first
taking the subvariety defined by the constraint $E_0 + \cdots + E_3 =
0$, and then taking the quotient of this subvariety by the $\SU(2)$
action generated by the constraint.   The latter step amounts to
considering tetrahedra only modulo rotations.  When nonempty, the
resulting symplectic manifold is generically 2-dimensional, as one would
expect, since the geometry of a tetrahedron modulo rotations has two
degrees of freedom after the areas of its faces have been fixed.  Applying
geometric quantization we obtain the Hilbert space $\Inv(j_0 \tensor
\cdots \tensor j_3)$.  

3) The space $\Inv(j_0 \tensor j_1 \tensor j_2 \tensor j_3)$
has a concrete description which is well-known from the quantum
mechanics of angular momentum.  We can think of an element of this space
as an intertwiner $\iota \maps j_0 \tensor j_1 \to j_2 \tensor j_3$.
Since any tensor product $j \tensor k$ decomposes as
\[        j \tensor k \iso |j-k|\, \oplus \, |j - k|+1 \, \oplus \,
\cdots \, \oplus \, j+k , \]
there is a basis of $\Inv(j_0 \tensor \cdots \tensor j_3)$ labeled by
spins $j_4$ satisfying 
\be    |j_0 - j_1| \le j_4 \le j_0 + j_1 , \qquad
 |j_2 - j_3| \le j_4 \le j_2 + j_3  \label{cond1} \ee
and
\be           j_0 + j_1 + j_4,\; j_2 + j_3 + j_4 \in \Z. \label{cond2} \ee

A more geometrical way to think of this is as follows.  Classically, 
the quantity
\[ A_{01} =
 {1\over 4}((E_0 + E_1)\cdot (E_0 + E_1))^{1/2} \]
is the area of a parallelogram formed by the midpoints of four edges of the
tetrahedron.  At the quantum level, the corresponding area operator is 
\[ \hat A_{01} =
 {1\over 4}((\hat E_0 + \hat E_1)\cdot (\hat E_0 + \hat E_1))^{1/2}. \]
Since the operators $\hat E_0^I + \hat E_1^I$ generate an $\so(3)$
action on the space $j_0 \tensor \cdots \tensor j_3$ and the operator
$\hat A_{01}$ commutes with the closure constraint, we can find a
basis of $\Inv(j_0 \tensor \cdots \tensor j_3)$ consisting of
eigenvectors of $\hat A_{01}$, with eigenvalues of the form ${1\over
4}\sqrt{j_4(j_4+1)}$.  This is just the basis described in the
previous paragraph.

More generally, if we split the four vertices of the tetrahedron into
two pairs $\{i,j\}$ and $\{k,l\}$, the quantity
\[ A_{ij} =
 {1\over 4}((E_i + E_j)\cdot (E_i + E_j))^{1/2} \]
is the area of the parallelogram whose vertices are the midpoints of the
four edges of the tetrahedron other than the edges $ij$ and $kl$.  At
the quantum level, we may define a corresponding {\it parallelogram area
operator}
\[ \hat A_{ij} =
 {1\over 4}((\hat E_i + \hat E_j)\cdot (\hat E_i + \hat E_j))^{1/2} \]
on the space $\Inv(j_0 \tensor \cdots \tensor j_3)$.  
Note that there is really just one parallelogram area operator for 
each splitting, since the closure constraint implies
\[  \hat A_{ij} = \hat A_{kl}  .\]
For any splitting, the eigenvectors of the corresponding parallelogram
area operator form a basis of  $\Inv(j_0 \tensor \cdots \tensor j_3)$ as
described above.  However the operators corresponding to different
splittings do not commute.  Thus different splittings give different
bases.  The transformations between these bases are given by $6j$
symbols in a well-known way.

As a consequence of all this, given a 3-manifold $S$ with a
triangulation $\Delta$ and a splitting of each tetrahedron in $\Delta$,
there is a basis of $L^2(\A_S/\G_S)$ given by all labelings of
triangles and tetrahedra in $\Delta$ by spins satisfying conditions like
those in equations (\ref{cond1}) and (\ref{cond2}) for each tetrahedron.  

The noncommutativity of parallelogram area operators can also be
understood in terms of the symplectic approach sketched in remark 2
above.  Classically, the geometry of a tetrahedron with faces of fixed
area has two degrees of freedom, but these degrees of freedom do not
Poisson-commute.  Thus states of the quantum tetrahedron with faces of
fixed area should be described by only one quantum number.  To see these
two degrees of freedom explicitly, note that while there are three
parallelogram areas $A_{01}, A_{02},$ and $A_{03}$, they satisfy one
equation:
\[     A_{01}^2 + A_{02}^2 + A_{03}^2 = A_0^2 + A_1^2 + A_2^2
+ A_3^2 .\]
Interestingly, this relation continues to hold at the quantum level.

4) It would be worthwhile trying to find formulas for other
geometrical observables as operators in the above context, such as the
total scalar curvature, the total scalar extrinsic curvature, and the
Hamiltonian constraint.  It would be especially interesting to compare
the commutation relations of such operators to those obtained by
Thiemann \cite{T2} in the real-analytic context.  


\section{Spin foams as quantum 4-geometries}   \label{4.geom}

We now turn to the question of how spin foams describe quantum
4-geometries.  Let $M$ be a compact piecewise-linear 4-manifold equipped
with a triangulation $\Theta$.  We also assume there is a principal
bundle $\tilde P \to M$ with structure group $\Spin(4)$.  Using this
bundle we can do lattice gauge theory over $M$ as in Section
\ref{lattice.gauge}.  In particular, we can consider a $\Spin(4)$ spin
foam in $M$ and try to understand it as equipping $M$ with a `quantum
4-geometry'.  We can also do this for any formal linear combination of
spin foams.  Actually, it seems that only formal linear combinations of
spin foams satisfying certain constraints have a fully satisfactory
interpretation as quantum 4-geometries.  Nonetheless we begin by fixing
a particular spin foam in $M$ and seeing how far we can get.   

One can understand quite a bit about the geometry of a 4-manifold by
studying the geometry of 3-manifolds embedded in it.  Thus we begin by
considering 3-dimensional submanifolds of $M$.  Suppose we have an
oriented 3-dimensional piecewise-linear submanifold $S \subseteq M$ with
a triangulation $\Delta \subseteq \Theta$.  Since the group $\Spin(4)$
is a product of two copies of $\SU(2)$, the bundle $\tilde P$ is a
product of `left-handed' and `right-handed' $\SU(2)$ bundles $P^+$ and
$P^-$.  This implies that the space $\A_S$ of connections on $\tilde
P|_S$ can be written as a product 
\[      \A_S = \A_S^+ \times \A_S^- \]
where $\A_S^\pm$ is the space of connections on $P^\pm|_S$, and similarly
\[     \A_S/\G_S = \A_S^+/\G_S^+ \times \A_S^-/\G_S^- \]
where $\G_S^\pm$ is the space of gauge transformations on $P^\pm|_S$.
Thus we have 
\[     L^2(\A_S/\G_S) \iso L^2(\A_S^+/\G_S^+) \tensor L^2(\A_S^-/\G_S^-) \]
where both the `left-handed' and `right-handed' factors in this tensor
product are spanned by $\SU(2)$ spin networks in the dual 1-skeleton of
$S$.  In particular, both factors are isomorphic.  In what follows we
will arbitrarily choose to favor $L^2(\A_S^+/\G_S^+)$ in certain
constructions, influenced by the Ashtekar's strategy of quantizing
gravity using the left-handed spin connection
\cite{Ashtekar,Ashtekar2,Ashtekar3}.  However, we show that in the end
the physics is independent of this choice.

As in Section 3 we can define various geometrically interesting
operators on the left-handed Hilbert space $L^2(\A_S^+/\G_S^+)$.  We can
tensor these with the identity on the right-handed Hilbert space to
obtain corresponding operators on $L^2(\A_S/\G_S)$.  This allows us to
interpret states in the latter space as quantum 3-geometries.  Moreover,
as described in Section 2, any spin foam $F$ in $M$ determines a state
$F|_S \in L^2(\A_S/\G_S)$.  Thus spin foams in $M$ determine quantum
3-geometries for $S$.

More concretely, note that a spin foam in $M$ is a labeling of each
2-cell in the dual complex $\Theta^\ast$ by an irreducible
representation of $\Spin(4)$ together with a labeling of each 1-cell by
an intertwining operator.  The 1-cells and 2-cells in the dual complex
correspond to tetrahedra and triangles in the triangulation $\Theta$,
respectively.  Moreover, any irreducible representation of $\Spin(4)$ is
of the form $j^+ \tensor j^-$ where $j^+$ and $j^-$ are representations
of the left-handed and right-handed copies of $\SU(2)$.  Thus if we pick
a splitting of each tetrahedron in $\Theta$, a spin foam in $M$ amounts
to a labeling of each triangle and each tetrahedron in $\Theta$ by a
pair of spins.

As explained above, given any triangle $f$ in the triangulation of $S$
there is an area operator $\hat A_f^+$ on $L^2(\A_S/\G_S)$ coming from
the area operator on the left-handed Hilbert space $L^2(\A_S^+/\G_S^+)$.
We define the expectation value of the area of $f$ in the spin foam $F$
to be
\[  \langle F, \hat A_f^+ F\rangle = \langle F|_S, A_f^+ F|_S \rangle .\]
Note that any triangle $f$ in $\Theta$ lies in some submanifold $S
\subseteq M$ of the form we are considering, and the above quantity is
independent of the choice of $S$.  Thus the expectation value of the
area of any triangle in $M$ is well-defined in the quantum 4-geometry
described by any spin foam.  The space $S$ serves only as a
disposable tool for studying the geometry of the spacetime $M$.

Note that we can easily extend the above formula to formal linear
combinations of spin foams.  This allows us to think of $A_f^+$ as an
operator on the space of formal linear combinations of spin foams.  More
generally, we can define an area operator $\hat A_\Sigma^+$ for any
2-dimensional submanifold $\Sigma \subseteq M$ built from triangles in
$\Theta$ by adding up the area operators for the triangles it contains.

Similarly, for every tetrahedron $T$ in the
triangulation of $S$ there is a 3-volume operator $\hat V_T^+$ on
$L^2(\A_S/\G_S)$, and we can define the expectation value of the 3-volume
of $T$ in the spin foam $F$ to be 
\[ \langle F, \hat V_T^+ F\rangle = 
\langle F|_S, \hat V_T^+ F|_S \rangle .\]
Again the right-hand side is actually independent of $S$, since it
depends only on the data by which $F$ labels the tetrahedron $T$ and its
faces.   As with the area of a triangle, we can also extend $V_T$ to an
operator on the space of  formal linear combinations of spin foams.  
More generally, we can define the volume operator $\hat V_S^+$ for any
3-dimensional submanifold $S \subseteq M$ built from tetrahedra in
$\Theta$ to be the sum of volume operators for the tetrahedra it
contains.

Now we turn to the question of when a linear combination of spin foams
defines a quantum 4-geometry for which each 4-simplex is `flat'.  For
this we consider the special case when our submanifold $S \subseteq M$
is the boundary of a single 4-simplex $X$ in the triangulation of $M$.
Any state in $L^2(\A_S/\G_S)$ describes a quantum 3-geometry for $S$,
but as explained in the Introduction, we only expect states satisfying
certain constraints to describe quantum 3-geometries for which $S$ can
be regarded as the boundary of a `flat' quantum 4-simplex.  Following
ideas of Barrett and Crane \cite{BC}, we now describe the subspace
\[                \X \subset L^2(\A_S/\G_S)  \]
of states with this property.  We call $\X$ the `Hilbert space of the 
quantum 4-simplex'.  We regard a linear combination of spin foams 
$\sum c_i F_i$ as describing a quantum 4-geometry for which $X$ is
flat when the state $\sum c_i F_i|_S$ lies in this subspace.  

To obtain the subspace $\X$, we begin by studying the geometry of an
ordinary classical 4-simplex in Euclidean $\R^4$.  Let $S$ be a
4-simplex with vertices $0,1,2,3,4$. We can think of a 4-simplex in
$\R^4$ with one vertex at the origin as an affine map from $S$ to $\R^4$
sending the vertex $0$ to the origin.  Such a map is given by fixing
vectors $e_1,e_2,e_3,e_4 \in \R^4$ corresponding to the edges $01$,
$02$, $03$ and $04$, as in Figure 2.  
This data amounts to a `cotetrad', a linear map
from the tangent space of the vertex $0$ of $S$ to $\R^4$.  

\vbox{
\medskip
\centerline{\epsfysize=2in\epsfbox{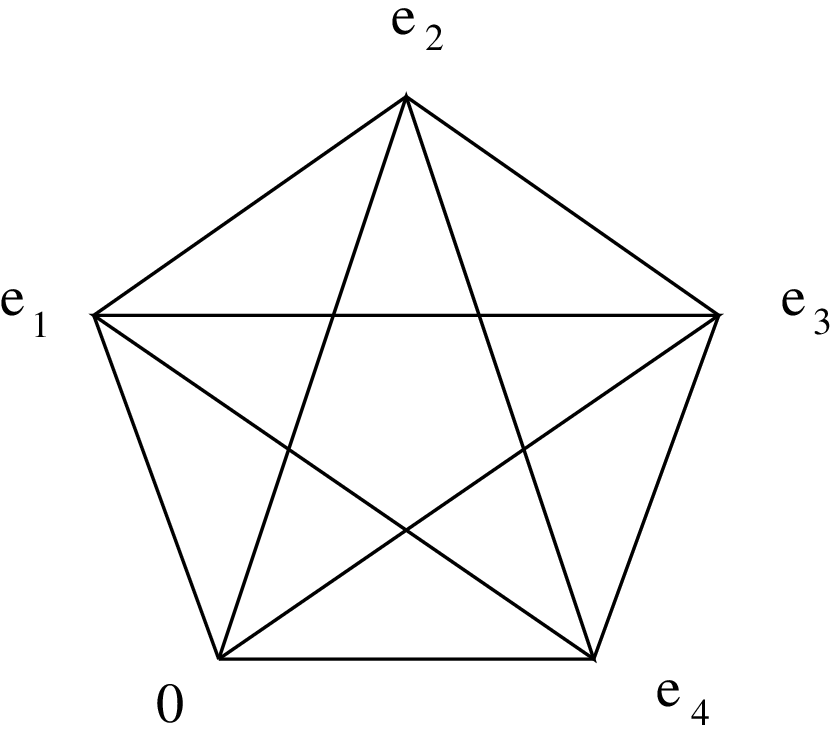}}
\bigskip
\centerline{2.  The 4-simplex as a cotetrad}
\medskip
}

Alternatively, we can describe the 4-simplex by associating bivectors to
its triangular faces.   For any pair $a,b = 1,2,3,4$ we can define a
bivector
\[          E_{ab} = e_a \we e_b  \]
corresponding to the face $0ab$.  However, not every collection of
bivectors $E_{ab}$ is of the form $e_a \we e_b$ for some vectors $e_a
\in \R^4$.  In addition to the obvious antisymmetry in the indices $a$
and $b$, the following constraints are also necessary:
\be      E_{ab} \we E_{cd} = 0 \; {\rm if} \;\{a,b\} \cap \{c,d\} \ne
\emptyset \label{type1} \ee
and 
\be       E_{12} \we E_{34} = E_{13} \we E_{42} = E_{14} \we E_{23}.
\label{type2} \ee

A naive count of degrees of freedom suggests that these conditions might
be sufficient: it takes 36 numbers to specify the bivectors $E_{ab}$, 
but there are 18 independent conditions of type (\ref{type1}) and 2 of
type (\ref{type2}), leaving us 16 degrees of freedom, exactly the number
of degrees of freedom in the vectors $e_a$.  Unfortunately, these
conditions are not quite enough.   In the generic case, when the
$E_{ab}$ ($a < b$) are linearly independent, it can be shown
\cite{Bryant,CDJ} that conditions (\ref{type1}) and (\ref{type2}) admit
exactly three sorts of solutions in addition to those of the form 
\[          E_{ab} = e_a \we e_b  \]
for some basis $e_a$ of $\R^4$.  The first is to take
\[          E_{ab} = -e_a \we e_b, \]
for some basis $e_a$, the second is to take
\[          E_{ab} = \sum_{a,b=1}^4 \epsilon_{abcd} e_c \we e_d, \]
and the third is to take 
\[          E_{ab} = - \sum_{a,b=1}^4 \epsilon_{abcd} e_c \we e_d . \]
If $E_{ab}$ can be written in any of these four ways, it can be written
so uniquely up to a parity transformation $e_a \mapsto -e_a$.  In what
follows, we simply ignore these subtleties and take any collection of
bivectors $E_{ab}$ antisymmetric in $a,b$ and satisfying (\ref{type1})
and (\ref{type2}) as an adequate substitute for a cotetrad.  A more
careful treatment would have to address these subtle problems, which are
evidently related to those already discussed for tetrahedra in three
dimensions.

Barrett and Crane make the all-important observation that conditions
(\ref{type1}) and (\ref{type2}) can be rewritten in a simpler way if we
use bivectors for all triangular faces of the 4-simplex, not just those
having $0$ as a vertex.  For this, it is convenient to set $e_0 = 0$ and
define
\[             E_{abc} = (e_c - e_b) \we (e_b - e_a) .\] 
The bivector $E_{abc}$ corresponds to the triangular face 
$abc$; in particular, we have $E_{0bc} = E_{bc}$.   The bivectors
$E_{abc}$ satisfy three sorts of constraints.  First, $E_{abc}$
is totally antisymmetric in the indices $a,b,c$.   Second, the 
bivectors $E_{abc}$ ($0 \le a < b < c \le 4$) satisfy five {\it closure
constraints} of the form: 
\be         E_{abc} - E_{abd} + E_{acd} - E_{bcd} = 0    
\qquad  \qquad a < b < c < d,   \label{closure} \ee 
one for each tetrahedral face of $S$.  Using these linear constraints,
conditions (\ref{type1}) and (\ref{type2}) can be rewritten as the {\it
quadratic constraints}
\be          E_{abc} \we E_{a'b'c'} = 0,  \label{quadratic} \ee
which hold whenever the triangles $abc$ and $a'b'c'$ share at least 
one edge --- that is, when they either share one edge or are the same.  
The quadratic constraints $E_{abc} \we E_{abc} = 0$ have a particularly
nice geometric interpretation: they say the bivectors $E_{abc}$ can be
written as wedge products of vectors in $\R^4$.  

Before attempting to quantize these constraints, we begin by quantizing
a single bivector in 4 dimensions, copying the procedure of 
Section 3.  Using a metric on $\R^4$ we identify $\Lambda^2 \R^4$ with
$\so(4)^\ast$, which as the dual of a Lie algebra can be thought of as a
classical phase space equipped with the Kirillov-Kostant Poisson
structure.  Since $\so(4)$ is isomorphic to $\so(3) \oplus \so(3)$, when
we quantize this phase space we obtain the {\it Hilbert space of a
quantum bivector in four dimensions},
\[   \H \tensor \H \iso \bigoplus_{j^+,j^-} j^+ \tensor j^- . \]
Note that this space is not a representation
of $\SO(4)$, but only of its universal cover, $\Spin(4)$.  

In what follows we write the Hilbert space of a quantum bivector in
four dimensions as $\H^+ \tensor \H^-$ to distinguish between the
`left-handed' and `right-handed' factors.  In fact, these correspond to
the self-dual and anti-self-dual parts of the quantum bivector, either
of which can be identified with a quantum bivector in three dimensions.
To see this, recall that given a metric and orientation on $\R^4$, we
may define the Hodge star operator
\[     \ast \maps \Lambda^2 \R^4 \to \Lambda^2 \R^4 . \] 
The $+1$ and $-1$ eigenspaces of this operator are called the spaces of 
self-dual and anti-self-dual bivectors, respectively: 
\[     \Lambda^2 \R^4 = \Lambda^2_+ \R^4 \oplus \Lambda^2_- \R^4 .\] 
This allows us to decompose a bivector $E$ in 4 dimensions into a
self-dual part $E^+$ and an anti-self-dual part $E^-$.  
Under the isomorphism between $\Lambda^2 \R^4$ and $\so(4)^*$, 
this splitting corresponds to the splitting
\[      \so(4)^\ast \iso \so(3)^\ast \oplus \so(3)^\ast. \]
Thus when we quantize $\so(4)^\ast$ with its Kirillov-Kostant
Poisson structure, we get a tensor product of Hilbert spaces
corresponding to the self-dual and anti-self-dual parts of
the space of bivectors.  

Now we turn to quantizing the constraints that say when ten bivectors
correspond to the faces of a flat 4-simplex.  We would like to 
interpret the closure constraints as imposing gauge-invariance.  The
closure constraints in equation (\ref{closure}) have minus signs, unlike
the closure constraint in our treatment of the quantum tetrahedron.  But
the reason for these signs is obvious: in our definition of the
$E_{abc}$, we implicitly chose orientations so that two triangular faces
of each tetrahedron are oriented clockwise and two are oriented
counterclockwise.  As in Section \ref{3.geom}, the solution we adopt
here is to explode $\Delta$ into a disjoint union of five tetrahedra,
thus doubling the number of triangles.  We can then define a bivector
for each face, orienting all the faces counterclockwise.  This amounts to
working with 20 bivectors $E_i(T)$, where $T$ ranges over the five
tetrahedra in $\Delta$ and $i = 0,1,2,3$ labels the faces of each
tetrahedron.  In these terms the closure constraints become 
\[ E_0(T) + E_1(T) + E_2(T) + E_3(T) = 0.  \]

To quantize the closure constraints this way, we start with
a Hilbert space
\[   \bigotimes_{T \in \Delta_3} (\H^+ \tensor \H^-)^{\otimes 4} \]
describing one quantum bivector for each face of the exploded complex.
On this space there are operators $E_i^{pq}(T)$ ($1 \le p,q \le 4$)
generating one $\so(4)$ action for each face $i = 0,1,2,3$ of each
tetrahedron $T$.  Then we pick out states $\psi$ for which
\[    (\hat E_0(T) + \hat E_1(T) + \hat E_2(T) + \hat E_3(T)) \psi  = 0 \]
for all $T$ in $\Delta$.  By the results of the previous section,
this gives the subspace 
\[        \bigotimes_{T \in \Delta_3} \T^+ \tensor \T^-  \]
where $\T^+$ and $\T^-$ are left-handed and right-handed copies
of the Hilbert space for the quantum tetrahedron introduced in Section 
\ref{3.geom}.  

Next, again motivated by Section 3, we impose left-handed and
right-handed versions of the equal-area constraints for pairs
of triangles in the exploded complex that came from the
same triangle in $\Delta$.   We use the splitting of $\so(4)$ 
into left-handed and right-handed copies of $\so(3)$ to write 
\[        \hat E_i(T) = \hat E_i^+(T) + \hat E_i^-(T),  \]
and define left-handed and right-handed area operators by 
\[    \hat A^\pm_i(T) =
 {1\over 2} (\hat E_i^\pm(T) \cdot \hat E_i^\pm(T))^{1/2}. \]
We then pick out states $\psi$ such that 
\[      A_i^+(T)\psi = A_{i'}^+(T')\psi \]
and 
\[      A_i^-(T)\psi = A_{i'}^-(T')\psi   \]
whenever $T,T'$ are tetrahedra in $\Delta$ with $\partial_i T 
= \partial_{i'} T'$.   This gives the smaller subspace
\[ \bigoplus_{\rho} \bigotimes_{T \in \Delta_3}  
\Inv(\rho_{\partial_0 T} \otimes \rho_{\partial_1 T} \otimes 
\rho_{\partial_2 T} \otimes \rho_{\partial_3 T}) \iso L^2(\A_S/\G_S) \]
where $\rho$ ranges over all labelings of triangles in $\Delta$
by representations of $\Spin(4)$, or in other words, pairs of spins.  

Finally we impose the quadratic constraints in the following
quantized form:
\be  (\hat E_i(T) \we \hat E_{i'}(T))\psi  = 0 \label{quantum.closure} \ee
States satisfying these constraints form {\it Hilbert space of the
quantum 4-simplex}, 
\[ \X = \{ \psi \in L^2(\A_S/\G_S) \; \colon \;
(\hat E_i(T) \we \hat E_{i'}(T))\psi  = 0 \; {\rm for \;
all}\; i,i',T \}  .\]
Barrett and Crane come close to explicitly describing this space.  The
key is to rewrite equation (\ref{quantum.closure}) as follows:
\[  (\hat E_i^+(T) \cdot \hat E_{i'}^+(T))\psi =
    (\hat E_i^-(T) \cdot \hat E_{i'}^-(T))\psi . \]

One can explicitly solve these constraints in the case $i = i'$.  Note
that a quantum bivector $\psi \in \H^+ \tensor \H^-$ satisfies
\[  (\hat E^+ \cdot \hat E^+)\psi = (\hat E^- \cdot \hat E^-)\psi   \]
if and only it lies in the subspace 
\[        \bigoplus_j j \tensor j  \subseteq \H^+ \tensor \H^-.  \]
By the same reasoning, the states in $L^2(\A_S/\G_S)$ satisfying
(\ref{quantum.closure}) for $i = i'$ are precisely those lying in
\[ \bigoplus_{\rho} \bigotimes_{T \in \Delta_3}  
\Inv(\rho_{\partial_0 T} \otimes \rho_{\partial_1 T} \otimes 
\rho_{\partial_2 T} \otimes \rho_{\partial_3 T}) \]
where $\rho$ ranges only over labelings of triangles in $\Delta$
by representations of the form $j \tensor j$.  

The case $i \ne i'$ is harder.  Note that for $i = i'$, we could also
write equation (\ref{quantum.closure}) in terms of the left-handed and
right-handed area operators as follows:
\[      \hat A_i^+(T) \psi =  \hat A_i^-(T) \psi  .\]
Similarly, we can express the quadratic constraints for $i \ne i'$ 
in terms of left-handed and right-handed versions of
the parallelogram area operators described in remark 3 of Section 3:
\[ \hat A_{ii'}^\pm(T) =
 {1\over 4}((\hat E_i^\pm(T) + \hat E_{i'}^\pm(T))\cdot 
 (\hat E_i^\pm(T) + \hat E_{i'}^\pm(T))^{1/2} .\]
Namely, having already imposed (\ref{quantum.closure}) for $i = i'$, the
constraint for $i \ne i'$ is equivalent to
\[      \hat A_{ii'}^+(T) \psi =  \hat A_{ii'}^-(T) \psi  .\]
The general solution of these equations is not yet known, 
but Barrett and Crane exhibit one solution for each
labeling of triangles of $\Delta$ by spins.  Thus the Hilbert
space $\X$ of a quantum 4-simplex is an infinite-dimensional subspace
of $L^2(\A_S/\G_S)$.  

Note that the quadratic constraints guarantee that all left-handed area
operators equal the corresponding right-handed operators when restricted
to $\X$.  In fact, one can show that vectors in $\X$ are invariant under
the transformation
\[       P \maps L^2(\A_S/\G_S) \to L^2(\A_S/\G_S)  \]
sending each vector $\psi^+ \tensor \psi^-$ in 
$L^2(\A_S^+/\G_S^+) \tensor L^2(\A_S^+/\G_S^-)$ to 
\[      P(\psi^+ \tensor \psi^-) = \psi^- \tensor \psi^+  .\]
Thus the left-handed and right-handed versions of all nonchiral
geometrical operators --- e.g., volume operators --- agree on states 
in $\X$.

We conclude with some miscellaneous remarks:

1) The spin foam description of quantum 4-geometries is a natural
outgrowth of work on `spinors and spacetime' \cite{Penrose-Rindler}.
This work applies not only to Riemannian geometry but also to Lorentzian
geometry.  It would thus be of great interest to study spin foams in the
Lorentzian context.  One will clearly need to use the well-known
relationships between $\SU(2)$, $\SL(2,\C)$ and $\SO(3,1)$.

2) It would be worthwhile to study the extent to which the role played
by simplices in this section and the last can be generalized to cells of
other shapes.  This is of physical interest because only very special
spin networks or spin foams lie in the complex dual to a triangulation.
In a physical theory, we should not impose this condition on our spin
networks and spin foams without a good physical reason.

3) The possibility of a purely left-handed description of quantum
4-geometries is of great interest, because most of the work relating
$BF$ theory to general relativity has focused on the left-handed
approach \cite{B4.5,CDJ,CDJM,Pleb,R3}.  For this it would be most
helpful to find a natural way to intepret $\X$ as a subspace of the
left-handed space $L^2(\A_S^+/\G_S^+)$.  This should be particularly
easy if the states of $\X$ found by Barrett and Crane are the only ones.
If such an interpretation exists, it might amount to a quantization of the
so-called `Capovilla-Dell-Jacobson' constraints which give general
relativity when added to $BF$ theory with gauge group $\SU(2)$.

4) Bivectors $E$ with $E \we E = 0$ are called {\it simple}; these are
precisely the bivectors that can be written as a wedge product of two
vectors in $\R^4$.  We may thus call $\bigoplus_j j \tensor j$ the {\it
Hilbert space of a simple quantum bivector}.  By the Peter-Weyl theorem,
this space is isomorphic to $L^2(\SU(2))$.  This `coincidence' deserves
to be better understood, because $L^2(\SU(2))$ plays a major role in the
spin network representation of general relativity formulated as a theory
of a left-handed spin connection \cite{AL2,B4}.  It is probably
important that both the proof that $\SU(2) \times \SU(2) = \Spin(4)$ and
the proof of the Peter-Weyl theorem rely on the left and right
actions of $\SU(2)$ on itself.

\section{Spin foam models} \label{model}

To actually extract new physical predictions about quantum gravity from
the spin foam formalism is likely to require new ideas and new tools
going far beyond those discussed in this paper.  In what follows we
merely give a taste of what this might be like.  In particular, we
sketch two spin foam models for 4-dimensional Euclidean quantum gravity.
The first model is just another way of looking at the state sum model of
Barrett and Crane \cite{BC}.  In this model one treats spacetime as a
4-manifold with a fixed triangulation and computes a discretized path
integral as a sum over all $\Spin(4)$ spin foams in the dual complex,
weighted by amplitudes computed as a product of face, edge, and vertex
amplitudes.  

The spacetime manifold plays a rather small role in this sort of theory.
It does not affect the computation of spin foam amplitudes; it only
constrains the set of spin foams being considered, via the requirement
that they live in the dual complex.  This suggests a second model where
spin foams are taken as fundamental and no reference is made to an
ambient manifold.  In this `abstract' spin foam model, spin foam
amplitudes are computed as before.  We simply drop the constraint that
the spin foams live in the dual complex of a fixed 4-manifold.  This
model is a crude attempt to combine Penrose's \cite{Moussouris,Penrose}
idea of building space from spin networks with Wheeler's \cite{W} vision
of spacetime as a kind of `foam' of vacuum fluctuations corresponding
roughly to manifolds with different topologies, but actually built from
some `pregeometry' of a more combinatorial nature.

\subsection*{Quantum gravity on a spacetime with fixed triangulation}

The state sum model of Barrett and Crane is based on the observation
that the Ooguri state sum model \cite{O} is a discrete formulation of
$BF$ theory, together with the observation that general relativity can
be viewed as $BF$ theory with extra constraints forcing the $E$ field to
be of the form $e \we e$ for some cotetrad field.  As we have seen, the
quantized version of these constraints is exactly what guarantees that a
spin foam represents a reasonable quantum 4-geometry --- i.e., one for
which every quantum 4-simplex is flat, and the left-handed and
right-handed 3-geometries of any submanifold coincide.  Barrett and Crane
obtain their state sum model by taking the Ooguri model and imposing
these constraints in their quantized form.  

The Ooguri model makes sense for any compact Lie group $G$ and any
compact oriented 4-manifold $M$ equipped with a principal $G$-bundle $P
\to M$ and a triangulation $\Theta$.  It can be viewed as a rule to
compute an amplitude for any spin foam $F = (\kappa,\rho,\iota)$ in $M$.
Here $\kappa$ is the 2-skeleton of the dual complex $\Theta^\ast$, and
in what follows we assume the labelings $\iota$ are taken from an
orthonormal basis of intertwiners at each vertex.  The amplitude is
computed as a product of amplitudes for all faces, edges and vertices of
the dual complex.

To compute the amplitude for a vertex $v$, note that the spin foam $F$
gives a spin network state $F|_S \in L^2(\A_S/\G_S)$ where $S$ is the
boundary of the 4-simplex in $\Theta$ containing $v$.  To emphasize the
dependence on $v$ we write $F|_S$ as $F_v$.  In fact, $F_v$ is a
continuous function on $\A_S/\G_S$.  There is a linear functional $\mu$
on the space of continuous functions on $\A_S/\G_S$ given by evaluation
on the equivalence class of the flat connection \cite{B4.5}.  We define
the amplitude of the vertex $v$ to be $\mu(F_v)$.  To compute the edge
amplitudes, note that the intertwiner $\iota_e$ can be attached to its
adjoint $\iota_e^\ast$ to form a spin network in the 3-sphere, which we
denote by $\iota_e \iota_e^\ast$.  We define the amplitude for the edge
$e$ to be $1/\mu(\iota_e \iota_e^\ast)$, where $\mu$ stands for
evaluation on the flat connection on the 3-sphere.  Finally, the
amplitude for each face $f$ is simply the dimension of the
representation $\rho_f$ labeling that face.

For more details on computing face, edge and vertex amplitudes, see the
work of Ooguri \cite{O} and also that of Crane, Kauffman and Yetter
\cite{CKY}.  We warn the reader that our normalizations differ from
theirs slightly, and our notation much more so.  To make the connection,
it is worth noting that for $\SU(2)$ spin networks on the 3-sphere,
evaluation on the flat connection is the same as the Penrose evaluation,
up to certain pesky but important signs.  In particular, what we call
$1/\mu(\iota_e \iota_e^\ast)$ is expressed by Crane, Kauffman and
Yetter in terms of two `$\theta$ evaluations', and what we call
$\mu(F_v)$ is expressed by them using a `$15j$ symbol'.

To obtain their model, Barrett and Crane take $G = \Spin(4)$ and use the
same face and edge amplitudes as in the Ooguri model, while modifying
the vertex amplitudes.  They take as their vertex amplitude the quantity
$\mu(P(F_v))$, where
\[      P \maps L^2(\A_S/\G_S) \to \X  \]
is the projection onto the Hilbert space of the quantum 4-simplex.
Since this projection operator appears at every vertex of the dual
complex, this modification of the vertex amplitudes has the effect of
picking out precisely those linear combinations of spin foams that
correspond to reasonable quantum 4-geometries.  In this model, the
partition function is given formally as a sum over spin foams in $M$:
\[       Z(M) = \sum_{F {\rm \; in \; } M} Z(F)  \]
where the amplitude $Z(F)$ of each spin foam $F = (\kappa,\rho,\iota)$
in $M$ is computed as follows:
\[      Z(F) = \prod_{f \in \kappa_2} \dim(\rho_f) 
\prod_{e \in \kappa_1} {1\over \mu(\iota_e \iota_e^\ast)} 
\prod_{v \in \kappa_0} \mu(P(F_v)).  \]
However, because there are usually infinitely many spin foams in $M$, 
this sum is likely to diverge.  Similarly, if one has a triangulated
4-dimensional cobordism $M \maps S \to S'$ equipped with a
$\Spin(4)$-bundle, one can formally define the time evolution operator
\[       Z(M) \maps L^2(\A_S/\G_S) \to L^2(\A_{S'}/\G_{S'})  \]
as a sum over all spin foams in $M$ of the corresponding spin foam
operators
\[        F  \maps L^2(\A_S/\G_S) \to L^2(\A_{S'}/\G_{S'})  \]
weighted by appropriately defined amplitudes.  But again, in practice
this sum is likely to diverge.  

The reason for these problems is that $\Spin(4)$ has infinitely many
irreducible representations.  Barrett and Crane also consider a
$q$-deformed variant of their theory.  Since quantum groups have only
finitely many irreducible representations when $q$ is a suitable root of
unity, one obtains well-defined partition functions and time evolution
operators this way.  While the physical meaning of this $q$-deformed
theory is not completely clear, there is an obvious guess.  Previously,
Crane and Yetter \cite{CY} constructed a $q$-deformed version of the
Ooguri model and showed that with suitable normalizations the partition
function is not only convergent but also independent of the choice of
triangulation of $M$.  The author has argued \cite{B2} that the
Crane-Yetter model is a discrete formulation of $BF$ theory with a
cosmological term, i.e., with the action given by
\[ \int_M \tr (E \we F + {\Lambda \over 12} E \we E), \] 
where the deformation parameter $q$ is a function of the cosmological
constant $\Lambda$.  Just as general relativity can be described as a
constrained $BF$ theory, so general relativity with nonzero cosmological
constant can be described as a constrained $BF$ theory with cosmological
term.  All this taken together suggests that the $q$-deformed version of
Barrett and Crane's state sum is a discretization of Euclidean quantum
gravity with cosmological constant.  In this regard it is also
interesting to note Major and Smolin's work on quantum gravity with
cosmological constant using $q$-deformed spin networks \cite{MS}, 
the degenerate solutions of general relativity with cosmological
constant obtained from $BF$ theory with cosmological term \cite{B5},
and the recent work of Markopoulou and Smolin \cite{MarkS}.

When contemplating these issues it is useful to keep in mind the
analogies with 3-dimensional quantum gravity.  In three dimensions one
has the Ponzano-Regge \cite{PR} model based on $\SU(2)$ and the
Turaev-Viro \cite{TV} model based on the quantum group $\SU_q(2)$.  The
former corresponds to Euclidean quantum gravity with $\Lambda = 0$, but
is divergent, while the latter is convergent and corresponds to
Euclidean quantum gravity with $\Lambda \ne 0$.  These models can be
formulated either in terms of triangulations or dually in terms of spin
foams having a `special spine' as their underlying 2-dimensional
complex.  Roughly speaking, a `special polyhedron' is a 2-dimensional
complex with only generic singularities, such as one sees in a foam of soap
bubbles, and a `special spine' of a compact 3-manifold $S$ is a
special polyhedron embedded in $S$ whose complement is an open ball.
For more on these matters, see the books by Turaev \cite{Turaev} and
Hog-Angeloni, Metzler and Sieradski \cite{HMS}.

To apply these analogies to the 4-dimensional case, it helps to note
that the Crane-Yetter model, based on triangulations, is equivalent to
Turaev's earlier construction using spin foams having a `skeleton' as
their underlying complex \cite{Turaev}.  A `skeleton' of a compact
4-manifold $M$ is a special polyhedron embedded in $M$ whose complement
is an open 4-dimensional handlebody.

\subsection*{Quantum gravity without a background spacetime}

As we have seen, the heart of Barrett and Crane's state sum
model is the following formula for the amplitude $Z(F)$ of a 
spin foam $F = (\kappa,\rho,\iota)$:
\[      Z(F) = \prod_{f \in \kappa_2} \dim(\rho_f) 
\prod_{e \in \kappa_1} {1\over \mu(\iota_e \iota_e^\ast)}
\prod_{v \in \kappa_0} \mu(P(F_v))  \]
Since this formula does not depend on the ambient spacetime manifold,
but only on the spin foam itself, we may consider it more generally as a
formula for evaluating the amplitude of {\it any} $\Spin(4)$ spin foam
with four faces meeting at each edge and with ten faces and five edges
meeting at each vertex in a pattern dual to that of a 4-simplex.  This
gives an `abstract' spin foam model, where there is no picture of
spacetime other than that provided by the spin foam itself.  Using the
formula described in Section 3, we may think of each face of $\kappa$ as
`carrying' a certain area, which it would give to any imagined surface
it intersected transversely.  Similarly, we may think of each edge as
carrying a certain volume.  It remains a challenge to determine, if
possible, the 4-volume associated to any vertex.  The main point,
however, is that if we think of an abstract spin foam as a kind of
quantum 4-geometry, the above formula serves as a rule for computing the
amplitude of any such quantum 4-geometry.

There are, of course, very serious problems in extracting physics
from such a model.  Most obviously, it is difficult to sum over
all spin foams.  Nonetheless it seems worth investigating such
models.  

\section*{Appendix A: Piecewise linear cell complexes}

We follow the definition of piecewise linear cell complexes given by
Rourke and Sanderson \cite{RoSa}.  A subset $X \subseteq \R^n$ is said 
to be a {\it polyhedron} if every point $x \in X$ has a neighborhood in 
$X$ of the form
\[       \{ \alpha x + \beta y \,\colon\; \alpha,\beta \ge 0, \;
\alpha + \beta = 1, \; y \in Y\}   \]
where $Y \subseteq X$ is compact.  A compact convex polyhedron $X$ for
which the smallest affine space containing $X$ is of dimension $k$ is
called a {\it $k$-cell}.  The term `polyhedron' may be somewhat
misleading to the uninitiated; for example, $\R^n$ is a polyhedron, and
any open subset of a polyhedron is a polyhedron.  Cells, on the
other hand, are more special.  For example, every 0-cell is a point, every
1-cell is a compact interval affinely embedded in $\R^n$, and every
2-cell is a convex compact polygon affinely embedded in $\R^n$.

The `vertices' and `faces' of a cell $X$ are defined as follows.  Given
a point $x \in X$, let $\langle x,X\rangle$ be the union of lines $L$
through $x$ such that $L \cap X$ is an interval with $x$ in its
interior.  If there are no such lines, we define $\langle x,X\rangle$ to
be $\{x\}$ and call $x$ a {\it vertex} of $X$.  One can show that
$\langle x,X\rangle \cap X$ is a cell, and such a cell is called a {\it
face} of $X$.

One can show that any cell $X$ has finitely many vertices $v_i$ and that
$X$ is the convex hull of these vertices, meaning that:
\[      X = \{ \sum \alpha_i v_i \, \colon\; 
\alpha_i \ge 0, \sum \alpha_i = 1\}  .\]
Similarly, any face of $X$ is the convex hull of some subset of the
vertices of $X$.  However, not every subset of the vertices of $X$ has a
face of $X$ as its convex hull.  If the cell $Y$ is a face of $X$ we
write $Y \le X$.  This relation is transitive, and if $Y,Y' \le X$ we
have $Y \cap Y' \le X$.

Finally, one defines a {\it piecewise linear cell complex}, or {\it
complex} for short, to be a collection $\kappa$ of cells in some
$\R^n$ such that: 
\begin{enumerate}
\item If $X \in \kappa$ and $Y \le X$ then $Y \in \kappa$.
\item If $X,Y \in \kappa$ then $X\cap Y \le X,Y$.  
\end{enumerate}
The union of the cells of a complex $\kappa$ is a polyhedron which
we denote by $|\kappa|$.  

A complex is {\it $k$-dimensional} if it has cells of dimension $k$ but
no higher.    A {\it subcomplex} of a complex $\kappa$ is a subset of
$\kappa$ which is again a complex, and  the {\it $k$-skeleton} of
$\kappa$ is the subcomplex consisting of all cells of dimension $k$ or
less.

\section*{Appendix B: The category of spin foams}

For each compact group $G$ we shall define a category $\cal F$ of spin
foams.  It is convenient here to restrict attention to `nondegenerate'
spin networks and spin foams.  A spin network is {\it nondegenerate} if
every vertex is the endpoint of at least one edge and every edge is
labeled with a nontrivial irreducible representation of $G$.  A spin
foam is {\it nondegenerate} if every vertex is the endpoint of at least
one edge, every edge is the edge of at least one face, and every face is
labeled with a nontrivial irreducible representation of $G$.  

The objects of the category $\cal F$ of spin foams are nondegenerate
spin networks with $G$ as gauge group.  As explained in Section 1,  to
define morphisms between arbitrary objects of $\cal F$ we need a
way to form a tensor product of spin networks whose underlying complexes
may not be disjoint, or may not lie in the same space $\R^n$.  Thus suppose
that $\Psi = (\gamma,\rho,\iota)$ and $\Psi' = (\gamma', \rho', \iota')$
are spin networks in $\R^n$ and $\R^{n'}$, respectively.  Suppose that
$k,k' > 0$ satisfy $n + k =  n' + k' = N$.  If we map $\gamma$ and
$\gamma'$ into $\R^N$ using the maps 
\[      (x_1,\dots,x_n) \mapsto (x_1,\dots,x_n,0,\dots,0) \]
and
\[       (x_1,\dots,x_n) \mapsto (x_1,\dots,x_{n'},0,\dots,1) \]
respectively, their images are disjoint.  We call the union of their
images a {\it disjoint union} of $\gamma$ and $\gamma'$.  We define a {\it
tensor product} $\Psi \tensor \Psi'$ to be a spin network whose 
underlying 1-dimensional oriented complex is a disjoint union of
$\gamma$ and $\gamma'$, with edges and vertices labeled by
representations and intertwiners using $\rho,\rho'$ and $\iota,\iota'$. 

We then define a morphism in $\cal F$ from $\Psi$ to $\Psi'$ to be a
certain equivalence class of nondegenerate spin foams $F \maps \emptyset
\to \Psi^\ast \tensor \Psi'$.  Two spin foams $F = (\kappa,\rho,\iota)$
and $F' = (\kappa',\rho',\iota')$  are regarded as {\it equivalent} if
one can be obtained from the other by a sequence of the following moves
and their inverses:

\begin{enumerate}
\item Affine transformation.  Suppose $F$ is a spin foam in $\R^n$ and
$F'$ is a spin foam in $\R^{n'}$.   Then $F'$ is obtained from $F$ by
{\it affine transformation} if: 

\begin{alphalist}
\item there is a one-to-one affine map $\phi \maps \R^n \to \R^{n'}$
establishing a one-to-one correspondence between cells of $\kappa$ and
cells of $\kappa$ and preserving the orientations of all cells;
\item if $f$ is a face of $\kappa$, then $\rho_f = \rho'_{\phi(f)}$;
\item if $e$ is a edge of $\kappa$, then $\iota_e = \iota'_{\phi(e)}$.
\end{alphalist}

\item Subdivision.  Suppose $F$ and $F'$ are spin foams in 
$\R^n$.   Then $F'$ is obtained from $F$ by {\it subdivision}
if: 

\begin{alphalist} 
\item the oriented 2-dimensional complex $\kappa'$ is obtained
from $\kappa$ by subdivision;
\item if the face $f'$ of $\kappa'$ is contained in the face $f$
of $\kappa$ then $\rho'_{f'} = \rho_f$;
\item if the edge $e'$ of $\kappa'$ is contained in the edge $e$
of $\kappa$ then $\iota'_{e'} = \iota_e$, while if $e'$ is the
edge of two faces of $\kappa'$ contained in the same face $f$ of
$\kappa$ then $\iota'_{e'} = 1_{\rho_f}$.
\end{alphalist}

\item Orientation reversal.  Suppose $F$ and $F'$ are spin foams in 
$\R^n$.  Then $F'$ is obtained from $F$ by {\it orientation reversal}
if: 

\begin{alphalist}
\item the 2-dimensional oriented complexes $\kappa$ and $\kappa'$ have the 
same cells, with possibly differing orientations;
\item if $f$ is a face of $\kappa$, then $\rho'_f = \rho_f$ if $f$
is given the same orientation in $\kappa$ and $\kappa'$, while $\rho'_f 
= (\rho_f)^\ast$ if it is given opposite orientations in $\kappa$ and
$\kappa'$;
\item if $e$ is a edge of $\kappa$, then $\iota_e$ equals $\iota'_{\phi(e)}$
after appropriate dualizations.
\end{alphalist}

\end{enumerate}

Here it is worth noting a subtle point.   A spin foam $F \maps \emptyset
\to \Psi^\ast \tensor \Psi'$ has been defined as a 2-dimensional
oriented complex $\kappa$ bordered by the disjoint union $\gamma \cup
\gamma'$, together with certain labelings of the edges and faces of
$\kappa$.   It is important to note that part of the data comprising $F$
is a one-to-one affine map from $|\gamma \cup \gamma'| \times [0,1]$ to
$|\kappa|$.  We use this `attaching map' to think of $\gamma$ and
$\gamma'$ as contained in $\kappa$.  When two spin foams are related by
affine transformation as in the definition above, their attaching maps
are required to be related by the same affine transformation in an
obvious way.  When two spin foams are related by subdivision or
orientation reversal, their attaching maps are required to be the same.

We define composition in $\F$ as follows.  Suppose we have morphisms
$F \maps \Psi \to \Psi'$ and $F' \maps \Psi' \to \Psi''$.  Then 
we can choose representatives of $F$ and $F'$ living in the same
space $\R^n$ such that the copy of $\Psi' = (\gamma', \rho', \iota')$ 
contained in $F$ is the same as the copy contained in $F'$, 
and the affine maps $c,c' \maps |\gamma'| \times [0,1] \to \R^n$
by which $\gamma$ borders the underlying complexes of $F$ and $F'$ fit
together to define a single affine map $f \maps |\gamma'| \times
[-1,1] \to \R^n$.  We then define the composite $FF'$  to be the spin
foam whose underlying complex is the union of those of $F$ and $F'$,
with faces and edges inheriting their labels from $F$ and $F'$, except
for the edges of $\gamma'$, which are labeled by appropriately dualized
forms of identity intertwiners.    

One can check that with this definition $\cal F$ becomes a category.
We conclude with two remarks:

1) The reader may wonder why we take affine transformation, subdivision
and orientation reversal as the basic forms of equivalence between spin
foams instead of just piecewise-linear homeomorphism and orientation
reversal.  Indeed, if we were only interested in abstract or
piecewise-linearly embedded spin foams, we could use such a simpler
notion of equivalence.  However, smoothly and real-analytically embedded
spin foams are also interesting, and one cannot define a smooth or
real-analytic embedding of a surface defined only up to piecewise-linear
homeomorphism!  The category we propose should be useful for studying
smoothly or real-analytically embedded spin foams, as well as
piecewise-linearly embedded ones.

2) Recall from remark 2 of Section 1 that, in addition to the `closed'
spin networks we have been using, there exist `open' spin networks. 
There should be a 2-category whose objects are finite collections of
points labeled by vectors in nontrivial irreducible representations of
$G$, whose morphisms are equivalence classes of open spin networks, and
whose 2-morphisms are equivalence classes of spin foams going between
such open spin networks.  Since there are tensor and duality operations
on spin networks and spin foams, we expect that this is a monoidal
2-category with duals --- in fact, a `strongly involutory' monoidal
2-category with duals.  This may be important in attempts to describe
quantum gravity in the language of higher-dimensional algebra
\cite{BD,Crane}.  However, we shall not endeavor to make this precise
here.  

\subsection*{Acknowledgments}  
I am grateful to the Center for Gravitational Physics and Geometry at
Pennsylvania State University for their hospitality while part of this
work was done, and especially thank Roumen Borissov,  Kirill Krasnov,
Fotini Markopoulou, Carlo Rovelli, Lee Smolin, and Jose Antonio Zapata
for many helpful discussions while I was there.   I also thank Allan
Adler, Robt Bryant, John Barrett, Allen Knutson, Justin Roberts, Michael
Reisenberger, and Stephen Sawin for help with 2-dimensional complexes, 
state sum models, and the quantum 4-simplex.  This work was partially
supported by NSF grant PHY95-14240.

\end{document}